\documentstyle[mnextra]{mn}
\seceqnum           % equations numbered by section
\input{epsf}
\def\etal{{\rm et al.} }
\def\kms  {\,{\rm km \, s^{-1}}}
\def\Mpc  {\,{\it h}^{-1}\, {\rm Mpc}}
\def\kpc  {\,{\it h}^{-1}\, {\rm kpc}}
\def\Msol {\,{\it h}^{-1}\, {\rm M_\odot}}
\def\Lsol {\,{\it h}^{-2}\, {\rm L_\odot}}
\def\MoverL {\,{\rm M_\odot/L_\odot}}
\def\d    {{\rm d}}

\def\ie {{\rm i.e. }}
\def\eg {{\rm e.g. }}
\def\log {{\rm log}}

\def\lsim{\mathrel{\hbox{\rlap{\hbox{\lower4pt\hbox{$\sim$}}}\hbox{$<$}}}}
\def\gsim{\mathrel{\hbox{\rlap{\hbox{\lower4pt\hbox{$\sim$}}}\hbox{$>$}}}}

\def\bj{b_{\rm J}}
\def\rf{r_{\rm F}}

\def\m@th{\mathsurround=0pt }
\def\eqalign#1{\null\,\vcenter{\openup1\jot \m@th
 \ialign{\strut\hfil$\displaystyle{##}$&$\displaystyle{{}##}$\hfil
 \crcr#1\crcr}}\,}

\begin{document}
\title[Luminous content of 2dFGRS galaxy groups]
{Galaxy groups in the 2dFGRS: the luminous content of the groups}
\author[V.R. Eke et al. (the 2dFGRS Team)]
{\parbox[t]\textwidth{
V.R. Eke$^{1}$,
Carlos S. Frenk$^1$,
Carlton M. Baugh$^{1}$,
Shaun Cole$^1$,
Peder Norberg$^{2}$,
John A. Peacock$^{3}$,
Ivan K. Baldry$^{4}$,
Joss Bland-Hawthorn$^5$,
Terry Bridges$^5$,
Russell Cannon$^5$,
Matthew Colless$^6$,
Chris Collins$^7$,
Warrick Couch$^8$,
Gavin Dalton$^{9,10}$,
Roberto de Propris$^8$,
Simon P. Driver$^6$,
George Efstathiou$^{11}$,
Richard S. Ellis$^{12}$,
Karl Glazebrook$^4$,
Carole A. Jackson$^6$,
Ofer Lahav$^{11}$,
Ian Lewis$^9$,
Stuart Lumsden$^{13}$,
Steve J. Maddox$^{14}$, 
Darren Madgwick$^{15}$,
Bruce A.\ Peterson$^6$,
Will Sutherland$^{10}$,
Keith Taylor$^{12}$ (the 2dFGRS Team)}
\vspace*{6pt} \\
{$^1$Department of Physics, University of Durham, South Road,
    Durham DH1 3LE, UK} \\
{$^2$ETHZ Institut f\"ur Astronomie, HPF G3.1, ETH H\"onggerberg, CH-8093
       Z\"urich, Switzerland} \\
{$^3$Institute for Astronomy, University of Edinburgh, Royal 
       Observatory, Blackford Hill, Edinburgh EH9 3HJ, UK}\\
{$^4$Department of Physics \& Astronomy, Johns Hopkins University,
       Baltimore, MD 21218-2686, USA} \\
{$^5$Anglo-Australian Observatory, P.O.\ Box 296, Epping, NSW 2121,
    Australia}\\
{$^6$Research School of Astronomy and Astrophysics,
 The Australian National University, Canberra, ACT 2611, Australia}\\
{$^7$Astrophysics Research Institute, Liverpool John Moores University,
    Twelve Quays House, Birkenhead, L14 1LD, UK} \\
{$^8$Department of Astrophysics, University of New South Wales, Sydney,
    NSW 2052, Australia} \\
{$^9$Department of Physics, University of Oxford, Keble Road,
    Oxford OX1 3RH, UK} \\
{$^{10}$Rutherford Appleton Laboratory, Chilton, Didcot, OX11 0QX} \\
{$^{11}$Institute of Astronomy, University of Cambridge, Madingley Road,
    Cambridge CB3 0HA, UK} \\
{$^{12}$Department of Astronomy, California Institute of Technology,
    Pasadena, CA 91125, USA} \\
{$^{13}$Department of Physics, University of Leeds, Woodhouse Lane,
       Leeds, LS2 9JT, UK} \\
{$^{14}$School of Physics \& Astronomy, University of Nottingham,
       Nottingham NG7 2RD, UK} \\
{$^{15}$Department of Astronomy, University of California, Berkeley, CA 92720, USA}\\ 
}
\maketitle
\begin{abstract}
The 2dFGRS Percolation-Inferred Galaxy Group (2PIGG) catalogue of
$\sim 29000$ objects is
used to study the luminous content of galaxy systems of various sizes.
Mock galaxy catalogues constructed from cosmological simulations 
are used to gauge the accuracy with 
which intrinsic group properties can be recovered.
It is found that a Schechter function is a
reasonable fit to the galaxy luminosity functions in groups of
different mass in the real data, and that
the characteristic luminosity $L_*$ is larger for more massive groups. 
However, the mock data show that the shape of the recovered luminosity
function is expected to differ from the true shape, and this
must be allowed for when interpreting the data.
Luminosity function results are presented in
both the $\bj$ and $\rf$ wavebands. The variation of halo
mass-to-light ratio, $\Upsilon$, with group size is studied in both these
wavebands. A robust trend of increasing $\Upsilon$ with increasing
group luminosity is found in the 2PIGG data. Going from groups
with $\bj$ luminosities equal to $10^{10}\Lsol$ to those $100$ times
more luminous, the typical $\bj$-band mass-to-light ratio increases by
a factor of $5$, whereas the $\rf$-band mass-to-light ratio grows by a factor
of $3.5$. These trends agree well with the predictions of the
simulations which also predict a minimum in the
mass-to-light ratio on a scale roughly corresponding to the Local
Group. Our data indicate that if such a minimum exists, 
then it must occur at $L\lsim 10^{10}\Lsol$, below the
range accurately probed by the 2PIGG catalogue. According to the mock data, 
the $\bj$ mass-to-light ratios of the largest groups
are expected to be approximately 1.1 times the global value.
Assuming that this correction applies
to the real data, the mean $\bj$ luminosity density
of the Universe yields an estimate of $\Omega_m=0.26 \pm 0.03$. Various
possible sources of systematic error are considered, with the conclusion
that these can affect the estimate of $\Omega_m$ by up to 20\%.
\end{abstract}

\section{Introduction}

The distribution of galaxy luminosities, along with their mass-to-light
ratios and spatial distribution, represent key observations
that should be explained by theories of galaxy formation. 
In the $\Lambda$CDM
model of structure formation, these broad brush empirical
characterisations of the galaxy population result from a complicated
interplay between the lumpy 
growth and coalescence of dark matter haloes and the radiative
cooling, star formation and feedback associated with baryons within
these clumps. 
Not only are the galaxies signposts to
locate the underlying overdensities in the dark matter distribution,
their properties also provide a systematic record of the processes that
have taken place in those haloes and their progenitors.
To develop a deeper understanding of the impact of these
various processes, it is helpful to break down the global constraints
referred to above and consider their variation with halo mass. 

The determination of the luminosity function of galaxies has received much 
observational effort (\eg Blanton \etal 2001, 2003; Cole \etal 2001;
Kochanek \etal 2001; Norberg \etal 2002), and recently there have been
suggestions that the distribution of galaxy luminosities varies
between rich cluster environments and low density regions
(\eg Zabludoff \& Mulchaey 2000; Christlein 2000; Balogh \etal 2001; de
Propris \etal 2003). 
Furthermore, some authors find evidence that
a Schechter function does not well describe the galaxy luminosity
function in groups and clusters because of an excess of bright
galaxies (\eg Smith, Driver \& Phillipps 1997; Trentham \& Tully 2002; 
Christlein \& Zabludoff 2003)
While these empirical results are intriguing, a systematic
study with a large set of homogenous groups would add welcome
weight to these findings. What should one expect to find? This
question has been addressed using semi-analytical
prescriptions (White \& Frenk 1991; Kauffmann \etal 1993; Diaferio
\etal 1999; Benson \etal 2003b). These techniques provide a physically
motivated {\em ab initio} procedure for calculating the properties of
galaxy populations, which can be compared with the real Universe in order to 
constrain the physical processes and their implementation within the models.
Both Diaferio \etal and Benson \etal find that the galaxy luminosity
functions within haloes of different mass are not well described by Schechter
functions. Instead, they have an excess in the abundance at the bright
end which arises from the different processes that are important for
the growth of the large central galaxy 
relative to the other galaxies in the haloes (the satellites). In
smaller haloes, where a larger 
fraction of the group luminosity is typically locked up in the central
galaxy, this produces a more 
prominent deviation from a Schechter function at the bright end of the
group's galaxy luminosity function. Furthermore, the luminosity at
which this `central galaxy bump' occurs, increases with
halo mass. Testing the predictions from these models represents an
opportunity to learn about aspects of galaxy formation.

The mass-to-light ratios ($\Upsilon$) of groups represent another
important clue, 
the light-to-mass ratio essentially reflecting the efficiency with
which stars are formed within haloes of different mass. In large
haloes this should be determined by the rate of gas cooling, whereas
in small haloes other factors that cause energy to be
injected into the halo gas become effective at disrupting the 
formation of stars. The prediction of semi-analytical models is that
mass should be converted most efficiently into optical light in haloes of mass
$\sim 10^{12} \Msol$ (White \& Frenk 1991; Kauffmann \etal 1999; 
Benson \etal 2000). One way in which this can be tested is to take
a theoretically motivated mass function and match haloes with galaxy
groups of the same abundance as inferred from an 
empirically determined group luminosity function. This
reveals the variation of the typical $\Upsilon$ with
mass. Qualitatively, Marinoni \& Hudson (2002) find a trend similar to
that predicted by semi-analytical models, namely a minimum $\Upsilon$
at intermediate masses of $\sim 10^{12} \Msol$. 
Observationally, there have
been many studies that have directly measured $\Upsilon$ for clusters
and large groups of galaxies in various wavebands ($R$ --
Carlberg \etal 1996, 2001; Tucker \etal 2000; 
$V$ -- Schaeffer \etal 1993; David, Jones \&
Forman 1995; Cirimele, Nesci \& Tr\`evese 1997; Hradecky \etal 2000; $B$
-- Ramella, Pisani \& Geller 1997; Adami \etal 1998; Girardi \etal 
2000, 2002; Sanderson \& Ponman 2003; Tully 2004). 
Some studies suggest that the
mass-to-light ratio is larger in bigger systems, whereas other studies
find no significant variation.

In addition to the direct relevance of the mass-to-light ratio and its
dependence on halo mass for studies of galaxy formation, $\Upsilon$ is
an important quantity to measure since it provides one of the traditional
ways to estimate the mean matter density of the Universe (\eg Carlberg
\etal 1996). This requires an assumption about the relation between
the mass-to-light ratio of the groups or clusters and the mean cosmic
value. Traditionally, it has been assumed that the value of $\Upsilon$
for the most massive clusters is representative of the universal mean,
although for a long time this assumption was challenged in biased
models of galaxy formation (Davis \etal 1985). At any rate, the value
of $\Upsilon$ for clusters, together with an appropriate assumption
about its universality and knowledge of the total galaxy luminosity
density allows a simple estimate of the mean mass density for the
Universe.

The construction of the 2dFGRS Percolation-Inferred Galaxy Group
(2PIGG) catalogue (Eke \etal 2004) from the 2
degree Field Galaxy Redshift Survey (2dFGRS: Colless \etal 2001, 2003)
facilitates a direct calculation of the galaxy content in a large
sample of homogenous groups. The purpose of this paper is to
report the results of a decomposition, by group size, of the
luminosity functions of galaxies within groups, and the group
mass-to-light ratio. Mock catalogues constructed from cosmological
simulations are employed as a guide to the
accuracy with which these quantities can be inferred from the 2PIGG
sample. This is a vital step in the comparison of a model prediction
with the observational results.

Section~\ref{sec:data} contains a brief description of the group
catalogue and a quantitative study of its reliability. In Section~\ref{sec:glf}
the luminosity functions of galaxies within groups is shown for
different halo masses. The variation of total halo mass-to-light ratio
with halo size is investigated in Section~\ref{sec:mol}.

\section{Inferred group properties}\label{sec:data}

\begin{figure}
\centering
\centerline{\epsfxsize=8.4cm \epsfbox{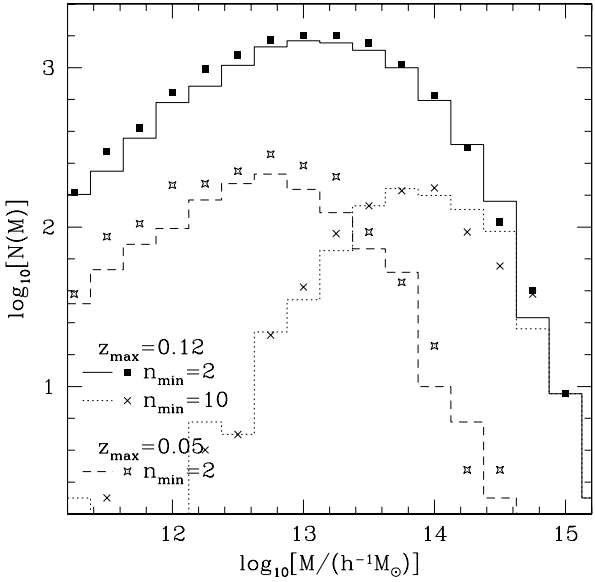}}
\caption{The number of groups at $z<0.12$ and $z<0.05$ 
as a function of the group
dynamical mass, $M$. Solid and dotted lines correspond to the
mock catalogues with $z_{\rm max}=0.12$ and $n_{\rm min}=2$ and $10$
respectively. Filled squares and crosses are likewise for the 2PIGG
catalogue. The dashed line is for $z_{\rm max}=0.05$ and $n_{\rm
min}=2$, and the corresponding result from the real catalogue is shown
with stars.}
\label{fig:nm}
\end{figure}

\begin{figure}
\centering
\centerline{\epsfxsize=8.4cm \epsfbox{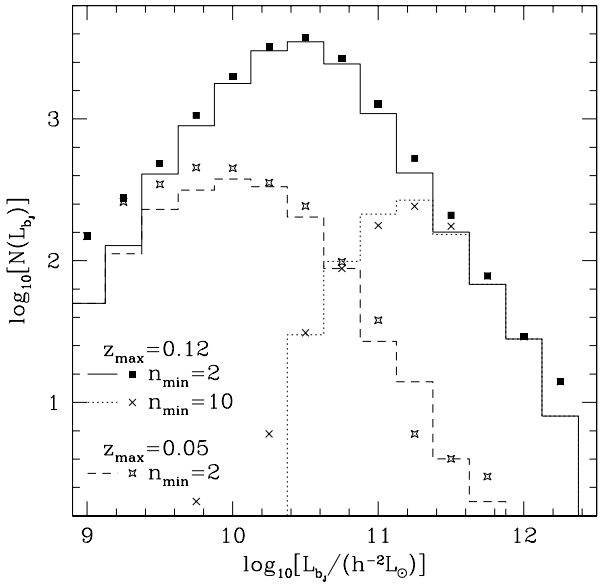}}
\caption{The number of groups as a function of the total
group $\bj$ band luminosity, $L_{b_{\rm J}}$. The lines and symbols have the same
meaning as in Fig.~\ref{fig:nm}.}
\label{fig:nl}
\end{figure}

A detailed description of the construction of the mock catalogues, and
the groupfinding algorithm applied both to these and the real 2dFGRS
was given by Eke \etal (2004). Briefly, three different $\Lambda$CDM dark
matter N-body simulations were used. They had ($L$, $N_p$,
$\sigma_8$) $=$ (154, $288^3$, 0.7), (250, $500^3$, 0.8) and (141,
$256^3$, 0.9), where $L$ is the side of the computational cube in $\Mpc$, $N_p$
the number of particles and $\sigma_8$ the rms linear density
fluctuation in 8$\Mpc$ spheres.
(The last of these is the GIF simulation described in
Jenkins \etal 1998.)  The semi-analytical model of galaxy
formation described by Cole \etal (2000) was implemented in the
simulations to create populations of model galaxies whose observable
properties are given by the model. 

Although the semi-analytical model produces a galaxy luminosity
function which is quite similar to that in the 2dFGRS, a
small rescaling of the $\bj$ luminosities was applied, to produce a model
luminosity function which was identical to that in the 2dFGRS.
Mock 2dFGRS catalogues were then constructed with the same
geometry and position-dependent flux limit as the 2dFGRS, and samples
of groups were selected by applying exactly the same cluster-finding
algorithm used to generate the 2PIGG catalogue. The $\rf$ band
magnitudes were rescaled in the same way as those in the $\bj$ band,
preserving the semi-analytical galaxy colours. Unlike in the
$\bj$ band, the resulting galaxy $\rf$-band luminosity functions for
the mock and 2dFGRS catalogues are thus not identical, as detailed in
Appendix~\ref{app:lr}.

Here, the focus is on describing the pertinent group properties and
quantifying the accuracy with which they are inferred. The latter
relies entirely on the mock catalogues, for which the true properties
of the haloes hosting the groups are known in the parent simulation.

As described by Eke \etal (2004), the group mass is inferred
dynamically according to
\begin{equation}
M=A\frac{\sigma^2 r}{G},
\label{mass}
\end{equation}
where $A=5.0$, $\sigma$ is the 1-dimensional velocity dispersion,
calculated using the gapper algorithm (Beers, Flynn \& Gebhardt 1990)
and removing $85 \kms$ in quadrature to account for redshift
measurement errors, and $r$ is the r.m.s. projected separation of
galaxies from the group centre, assuming an $\Omega_m=0.3,
\Omega_\Lambda=0.7$ cosmological model. The value of $A$ has been
chosen by reference to mock catalogues in which galaxies trace the
dark matter within each halo, apart from a `central' galaxy placed at
the halo centre of mass. There is some evidence from gravitational
lensing studies that bright galaxies do indeed trace the mass, at
least in rich clusters (Hoekstra \etal 2002; Kneib \etal 2003). 
However, these current observational constraints are not yet
sufficiently stringent that the effective value of $A$ can be
empirically justified to better than a few tens of per cent. This is
the main potential source of systematic error in this paper.

\begin{figure*}
\centering
\centerline{\epsfxsize=13.5cm \epsfbox{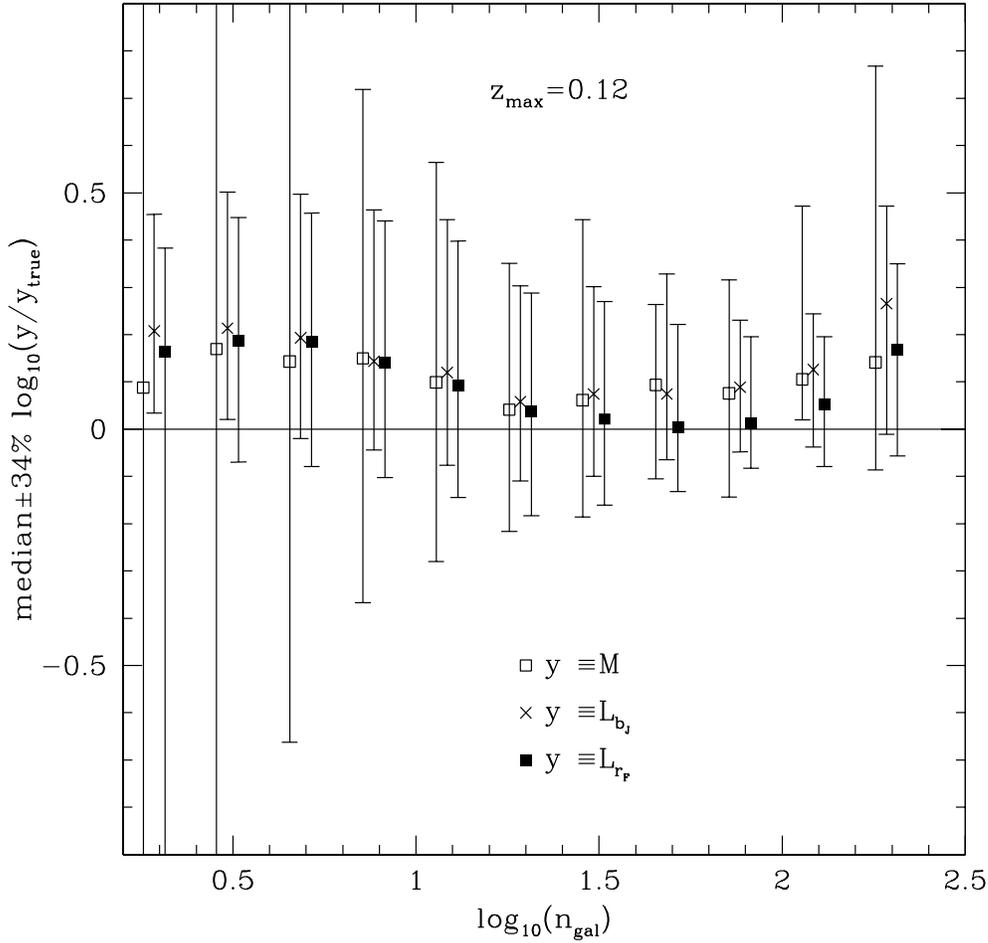}}
\caption{The median accuracies (symbols) $\pm 34$ percentiles (error bars) 
of the inferred group properties as a function of the number of
galaxies in a group, $n_{\rm gal}$. All groups with $z\le0.12$ are
included. The symbols correspond to the following group properties:
$y=M$ (open squares), $L_{b_{\rm J}}$ (crosses) and $L_{r_{\rm F}}$
(filled squares). The points have been slightly displaced either side
of the true bin values for clarity. }
\label{fig:errorsn}
\end{figure*}

\begin{figure*}
\centering
\centerline{\epsfxsize=13.5cm \epsfbox{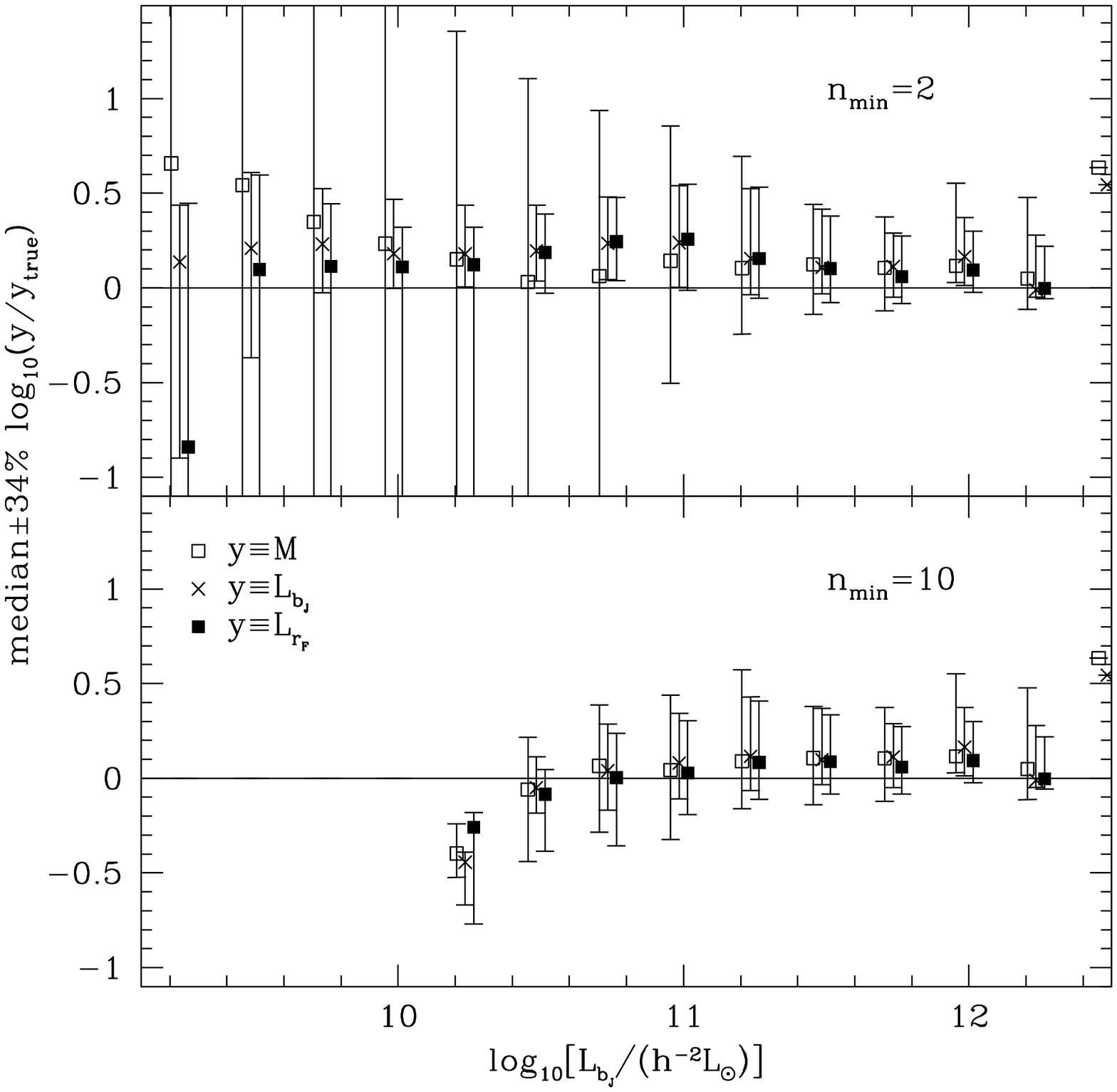}}
\caption{The median
accuracies (symbols) $\pm 34$ percentiles of the inferred group
properties as a function of the inferred group luminosity, $L_{b_{\rm
J}}$. In the top panel, $n_{\rm min}=2$ and, in the bottom panel,
$n_{\rm min}=10$. Each bundle of three symbols has the following order
from left to right: $y=M, L_{b_{\rm J}}$ and $L_{r_{\rm F}}$. The
points have been slightly displaced either side of the true bin values
for clarity.}
\label{fig:errorsl}
\end{figure*}

\subsection{Calculating group luminosities}\label{ssec:lb}

In order to calculate the total group luminosity, it is necessary to
correct for the incompleteness in the 2dFGRS. When finding the groups,
galaxies from the parent catalogue that have no measured redshift have
their weights redistributed equally to the nearest 10 projected
galaxies with measured redshifts, and these galaxies are assigned a
larger linking volume accordingly. As the sets of galaxies with and
without redshifts are random subsamples (in terms of the intrinsic
galaxy properties) of the same underlying galaxy population (ignoring
the small level of flux-dependent redshift incompleteness), the
incompleteness can simply be taken into account by totting up the
observed group $\bj$ luminosity according to
\begin{equation}
L_{obs,\bj}=\Sigma_i^{n_{\rm gal}} w_i L_{i,\bj},
\label{easyw}
\end{equation}
where the sum extends over the $n_{\rm gal}$ galaxies in the group with their
individual weights, $w_i$.
The total group luminosity is then estimated by including 
the contribution from galaxies below the luminosity
limit, $L_{\rm min}$, at the redshift of the group. 
The extrapolation to zero galaxy luminosity is performed assuming that
a Schechter function describes the number of galaxies as a function of
luminosity. Given that the Schechter function is
\begin{equation}
\phi(L)\d L=\phi_* \left(\frac{L}{L_*}\right)^\alpha {\rm exp}\left({-\frac{L}{L_*}}\right)
\frac{\d L}{L_*},
\label{sch}
\end{equation}
this operation merely involves dividing the observed luminosity by
the incomplete Gamma function 
$\Gamma(\alpha+2,L_{\rm min}/L_*)/\Gamma(\alpha+2)$. In 
practice, a small correction is also applied because galaxies with
fluxes corresponding to $\bj<14$ have been removed from the redshift 
catalogue. This bright flux limit only makes a perceptible difference
for a few very local groups. These calculations require the
additional information that $M_\odot=5.33$ in the $\bj$ band and the
adopted $k+e$-correction is similar to that of
Norberg \etal (2002), 
\begin{equation}
k+e=\frac{z+6z^2}{1+8.9z^{2.5}}.
\label{kpluseb}
\end{equation}
The global $(M_*,\alpha)=(-19.725,-1.18)$ values are used to
extrapolate to the total luminosity for all groups. 
These values differ slightly from those of Norberg
\etal (2002) as they have been derived from the recalibrated 2dFGRS
(Colless \etal 2003).  Using a halo mass-dependent extrapolation (see
Section~\ref{sec:glf}) changes the inferred luminosities by no more
than $\sim 10$ per cent.

At $z>0.12$, the fraction of the total group luminosity that is
actually observed
drops below a half. Additionally, according to the mock catalogues,
the amount of contamination of group membership increases at these
higher redshifts. Thus, in all of what follows, only
groups at $z<0.12$ will be considered. This provides a large number of
well-sampled groups with a range of masses up to $\sim
10^{15}\Msol$. The number of groups as a function of mass and
luminosity is shown in Figs~\ref{fig:nm} and ~\ref{fig:nl} for both
the mock and real catalogues, choosing the minimum group membership to
be $n_{\rm min}=2$ or $10$. Distributions of $n_{\rm min}=2$
groups at $z<0.05$ are also shown in these figures. The mock and real
catalogues yield broadly similar numbers of groups with similar masses
and luminosities. There is a slight deficit of groups in the mock
catalogue relative to the 2PIGG data. This originates from the lack of
low luminosity galaxies included at $z<0.04$, as described by Eke \etal (2004).

Complications arise when using the SuperCOSMOS (Hambly \etal 2001)
$\rf$ band data because the 2dFGRS is a $\bj$ selected
survey. Different depths in the red luminosity function are thus
probed for galaxies of different spectral types. Rather than dealing 
with the explicit dependence of the luminosity function on colour, the
sample was re-cut to a conservative $\rf$ limit such that almost all
2dFGRS galaxies can be detected to the same red luminosity at a given
redshift.  This issue is discussed in detail in Appendix~A, which
justifies re-cutting the sample at a local limit of $r_{\rm
F,lim}=b_{\rm J,lim}-1.5$.

\subsection{The accuracy of measured halo properties}\label{ssec:accu}

The typical redshift error in the 2dFGRS of $\sim 85 \kms$ (Colless \etal
2001) will lead to small mass haloes having considerable errors in
their estimated velocity dispersions and hence masses.
For instance, $\sigma=85 \kms$ and $r=100 \kpc$
correspond to a mass of about $10^{12} \Msol$. In order to interpret
the results in the following sections, it is important to quantify the
degree of uncertainty in the estimated halo masses and luminosities and
how this varies with properties such as the number of member galaxies, 
$n_{\rm gal}$, or halo luminosity. 

The `accuracy' with which a halo property is measured is defined as
a logarithmic measure of the error in the property:
\begin{equation}
{\rm accuracy} \equiv \log_{10} \left(
{y_{\rm recovered} \over y_{\rm true}} \right),
\end{equation}
where $y$ can stand for mass, $\bj$ or $\rf$ luminosity.
Figs~\ref{fig:errorsn} and~\ref{fig:errorsl} show how the median
accuracies, and the 16th and 84th percentiles (as opposed to the mean
$\pm 1\sigma$), vary as functions of $n_{\rm gal}$ and $L_{b_{\rm J}}$
respectively. Fig.~\ref{fig:errorsn}
shows that the spread in mass accuracies becomes very large when
the number of group members is less than $10$. For more populated
groups, the median accuracy is approximately zero, meaning that the
typical inferred mass is unbiased. The luminosities are
typically overestimated by a few tens of per cent
even for well-populated groups, but they can still
be relatively well determined for binaries. From this figure, one can
infer that measured group luminosity is a better quantity than the
measured dynamical mass for ranking the groups in order of size.
Note that the accuracy of $\rf$ luminosities has a larger spread 
than that in the $\bj$ band. This is the result of using fewer
galaxies to calculate the $\rf$ luminosity, meaning
that the Schechter function correction for unobserved galaxies
is slightly larger in the $\rf$ than in the $\bj$ band.
Similar trends are visible in Fig.~\ref{fig:errorsl},
where the distributions of accuracies are shown as a
function of group luminosity for two different values of $n_{\rm
min}$. At low group luminosity, particularly for $n_{\rm min}=2$, the
halo masses become increasingly 
overestimated and poorly determined due to the impact of the 2dFGRS
redshift errors and contamination. Also apparent is a
tail of underestimated group $\rf$ luminosities at low $\bj$ 
group luminosity. These are binary groups with no members that
satisfy the adopted flux limits, and are thus assigned
zero $\rf$-band luminosity.

Armed with this quantitative understanding of the group catalogue, it
is now appropriate to see what information can be extracted concerning
the galaxy populations in different sizes of halo.

\section{Galaxy luminosity functions within groups}\label{sec:glf}

\begin{figure*}
\centering
\centerline{\epsfxsize=19cm \epsfbox{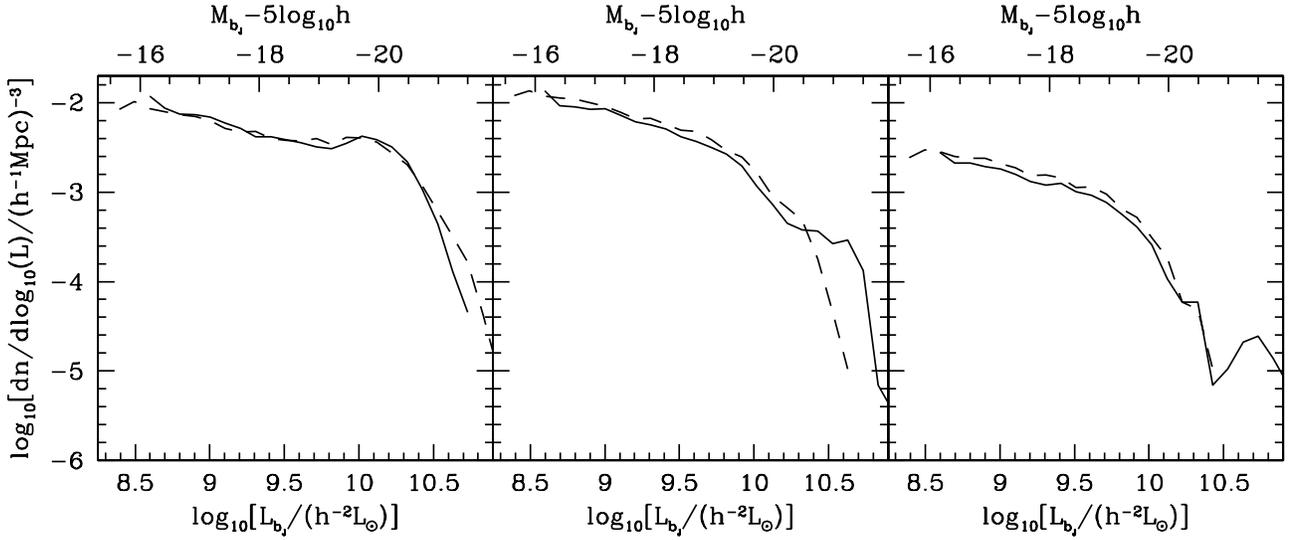}}
\vspace{-11.5cm}
\caption{Galaxy $\bj$ luminosity functions in haloes of different
mass predicted by two semi-analytical models of galaxy formation. The
panels refer to haloes with masses of $10^x\Msol$, with $x=13\pm0.5$
(left), $14\pm0.5$ (centre) and $15\pm0.5$ (right). The scales at the
top of the panels give the absolute $\bj$ magnitude. Solid lines
correspond to the `bumpy' semi-analytical model, whereas dashed
lines correspond to the `superwind' model.}
\label{fig:bcomp}
\end{figure*}

\begin{figure*}
\centering
\centerline{\epsfxsize=19.cm \epsfbox{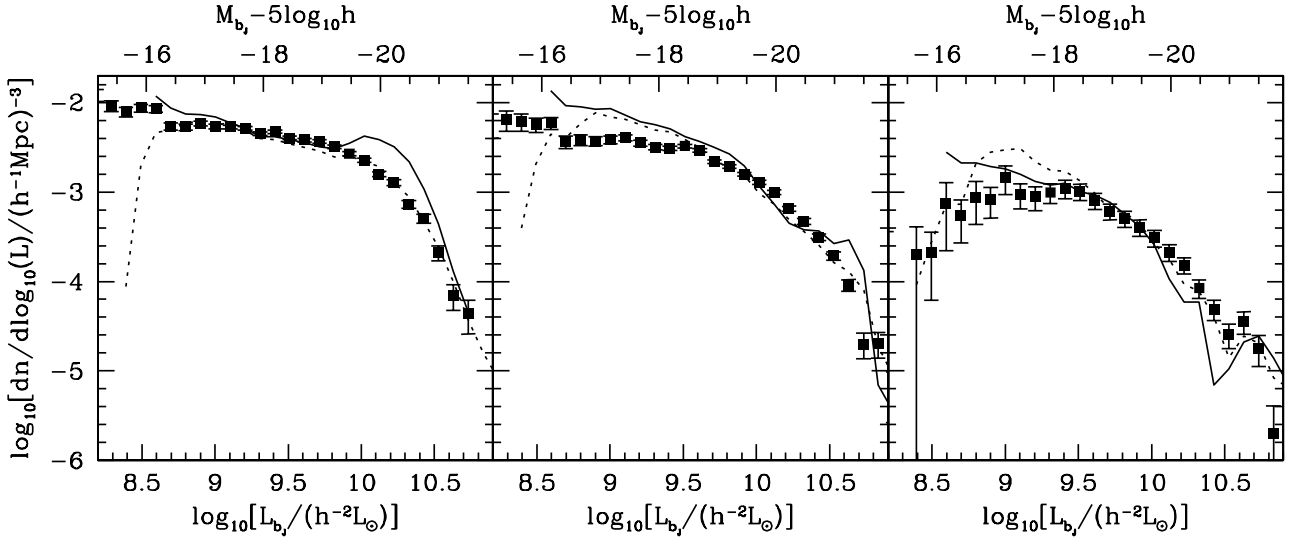}}
\vspace{-11.5cm}
\caption{Galaxy $\bj$ luminosity functions in groups of different mass
in both mock and real 2PIGG catalogues. The group mass ranges are
$M\sim 10^{13}\Msol$ (left), $M\sim 10^{14}\Msol$ (centre) and $M\sim
10^{15}\Msol$ (right). In each case, a solid line shows the
semi-analytical prediction from which the mock catalogues were
constructed, a dotted line shows what is actually measured in the mock
groups, and the filled squares with errors depict the results from the
2PIGG data. The values of $n_{\rm min}$ and $z_{\rm max}$ are selected
for each mass range, as described in the text, in order to optimize
the recovery of the original galaxy semi-analytical luminosity
functions.}
\label{fig:bbandb}
\end{figure*}

This Section addresses firstly the issue of how well the galaxy luminosity
function in haloes of different mass can be recovered from the mock
catalogues and, secondly, the determination of these functions for the
real data. In calculating the space
density of galaxies, a $1/V_{\rm max}$ estimator has been
applied to each galaxy. The variable flux limit across the survey has
been taken into account. Dynamically inferred group masses, calculated
according to equation~(\ref{mass}), have been used to split the total
sample into different classes.

\subsection{$b_{\rm J}$ band results}\label{ssec:glfb}

Sets of mock catalogues have been constructed from two different
semi-analytical models of galaxy formation applied to the same N-body
simulations. Both these models are based on the general scheme
developed by Cole \etal (2000), but they treat certain physical
processes in different ways. The first model, described by Benson
\etal (2002), is very similar to the original Cole \etal model except
that it includes detailed treatments of photoionization of the
intergalactic medium at high redshift and of the dynamics of
satellites in haloes. This model, which will be referred to as the
`bumpy' model for reasons that will become 
apparent shortly, was used by
Benson \etal (2003b) to predict the luminosity function of galaxies in
haloes of different mass. The second model comes from Benson \etal
(2003a) and will be referred to as the `superwind' model. This has a
baryon fraction 
twice as high as that assumed by Cole et al., in accordance with 
recent determinations, and includes a treatment of superwinds. These
suppress the growth of the overly bright galaxies which otherwise tend
to form in high baryon fraction models.

Fig.~\ref{fig:bcomp} shows galaxy luminosity functions in groups of
different mass in both the semi-analytical models of galaxy
formation. The solid lines correspond to the bumpy model while the
dashed lines correspond to the superwind model. The suppression of
bright central galaxies in the superwind model is apparent in the
middle and right-hand panels which show the galaxy luminosity
functions within groups of mass $M\sim 10^{14}\Msol$ and $M\sim
10^{15}\Msol$ respectively.
%% The following seems unnecessary and probably confusing -- csf
%When creating the mock catalogue, the galaxy luminosities are rescaled
%slightly in order to reproduce the overall 2dFGRS luminosity
%function. As a result, the overly so these brightest galaxies must
%exist somewhere. The left panel of the figure shows that at least some
%of them have been moved into the haloes with $M\sim
%10^{13}\Msol$. 
It is clear that the detailed shapes of the galaxy luminosity
functions in groups depend upon the assumptions that go into making a
semi-analytical model. These data can thus be used to test the
models.
%Both mock catalogues have been used to determine how well these
%functions are recovered. 
In what follows, the bumpy mock catalogues will be shown unless
otherwise stated.

In order to recover faithfully both the amplitude and shape of the
galaxy luminosity function in groups of different mass, it is helpful
to adopt separate values of $n_{\rm min}$ and $z_{\rm max}$ for the
various group samples. For instance, clusters that should contribute
to the highest mass bin will have large numbers of members and be
detectable to higher redshifts compared to the smaller groups. Thus,
the following empirical choices of ($n_{\rm min}$, $z_{\rm max}$) were
made for the $10^{13}$, $10^{14}$ and $10^{15}\Msol$ samples
respectively: $[(3,0.08), (10,0.12), (80,0.12)]$. These will be used
throughout this Section. 

The results of these selections can be seen in Fig.~\ref{fig:bbandb},
which shows the semi-analytical model luminosity functions in the
simulations from which the mock catalogues were created, the
luminosity functions actually recovered from the mocks, and the
luminosity functions estimated from the 2PIGG sample for the same
values of $n_{\rm min}$ and $z_{\rm max}$.  While the theoretical
luminosity functions in the simulations from which the mock catalogues
were made differ substantially from Schechter functions, the bumps
have been largely smeared out in the mock recovered luminosity
functions, which now better resemble Schechter functions. This
smearing is predominantly the result of contaminating galaxies
contributing to haloes with inappropriate masses and homogenizing the
samples. Note that the original bumps occur at luminosities that vary
with halo mass. There is a tentative bump detection in the two higher
mass group samples. This excess of luminous galaxies is sufficient to
render the Schechter function a bad fit, although it does provide a
good description of the results for the $M\sim 10^{13}\Msol$
groups. Overall however, apart from the smearing out of the central
galaxy bumps, the shapes and amplitudes of the group galaxy luminosity
functions are recovered well from the mock catalogue.

\begin{figure}
\centering
\centerline{\epsfxsize=8.4cm \epsfbox{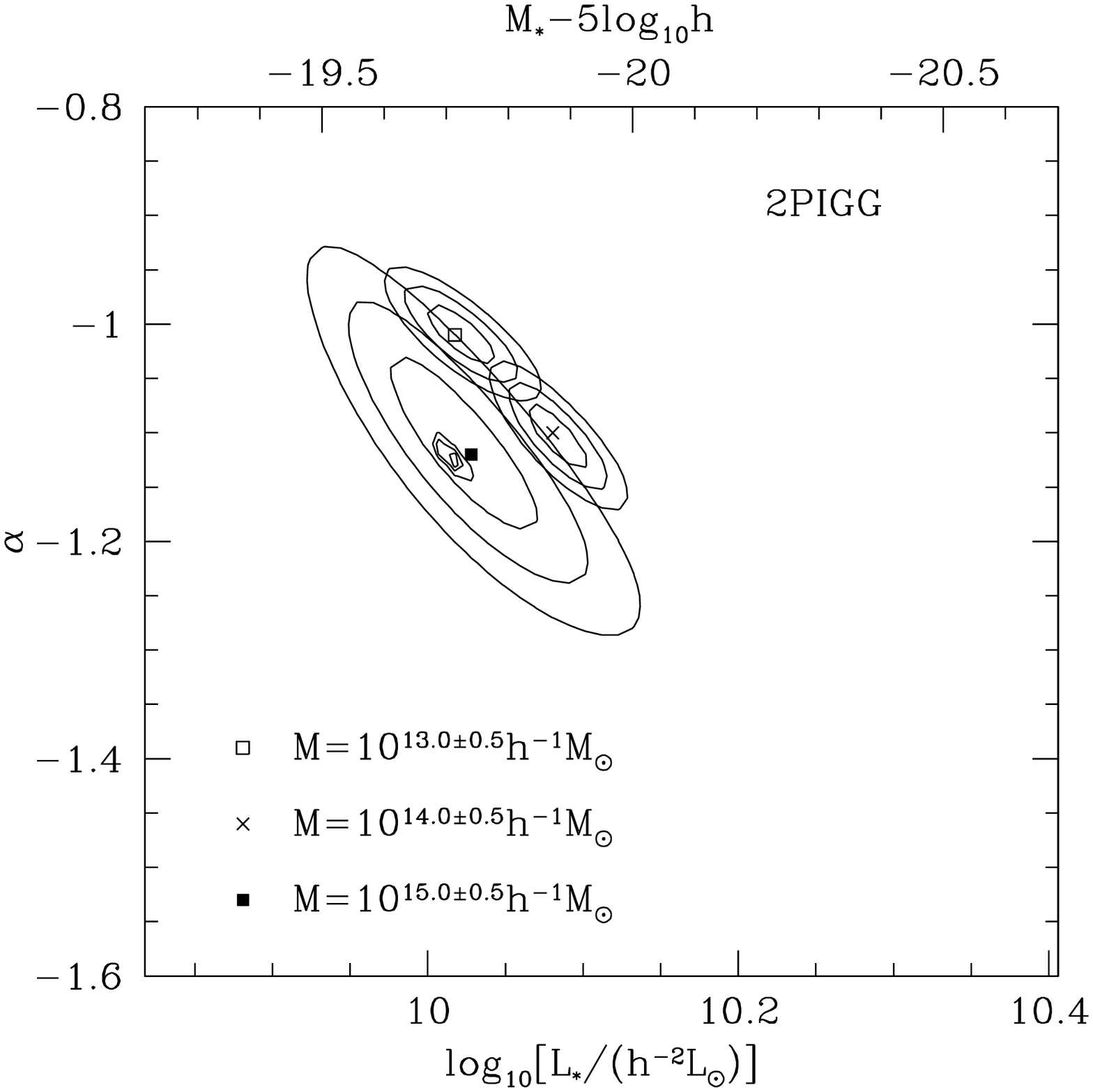}}
\caption{$1$, $2$ and $3-\sigma$ contours showing the relative
probability that different Schechter function parameters, $L_*$ and
$\alpha$, provide a
good description of the $\bj$-band galaxy luminosity function within
2PIGGs of different mass. These results were obtained by the STY
estimation method using the groups contributing to the 2PIGG luminosity functions
in Fig.~\ref{fig:bbandb}. The different group masses have different
symbols marking the most likely parameter values, as detailed in the
figure. The small ellipses with no central symbol represent the
results for all 2dFGRS galaxies at $z<0.12$.}
\label{fig:styb}
\end{figure}

\begin{figure*}
\centering
\centerline{\epsfxsize=19.cm \epsfbox{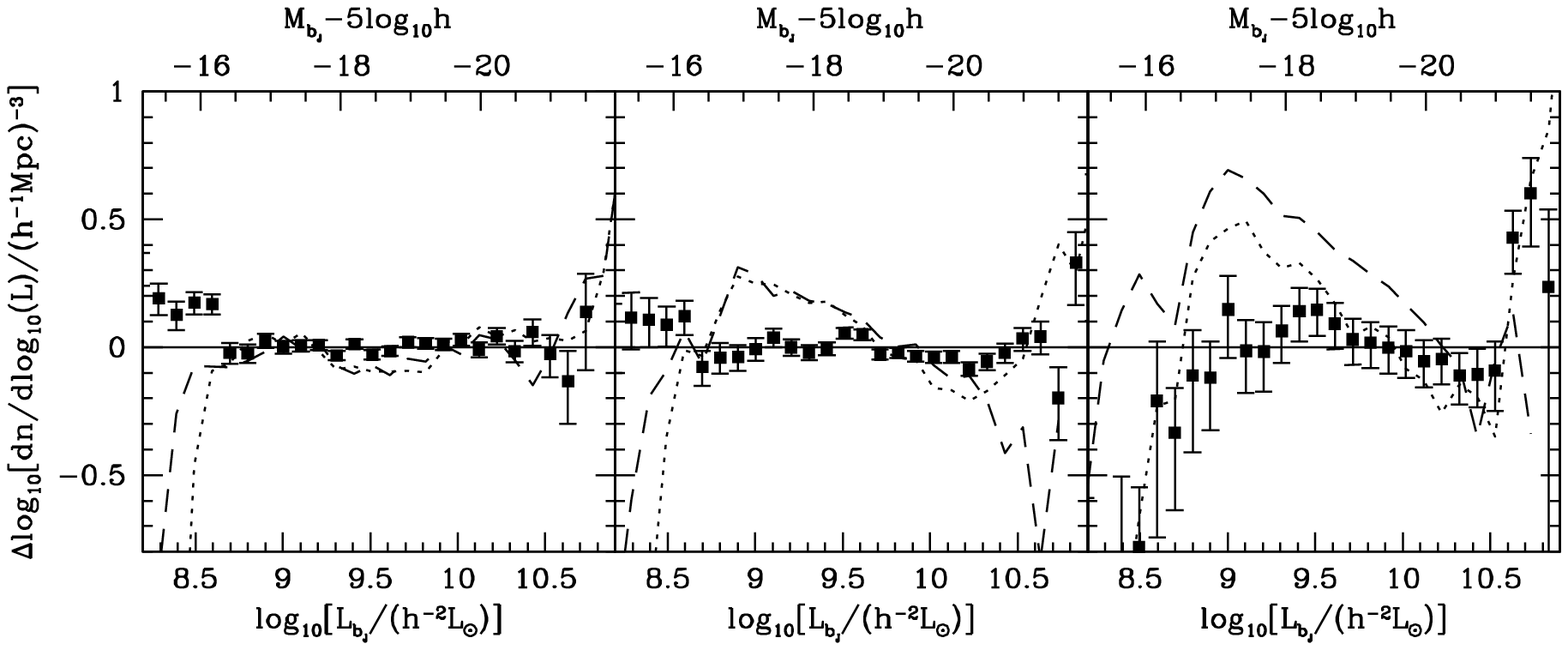}}
\vspace{-11.5cm}
\caption{The ratios of galaxy luminosity functions to
the 2PIGG best-fitting Schechter functions for the $M\sim 10^{13}\Msol$
(left), $M\sim 10^{14}\Msol$ (centre) and $M\sim 10^{15}\Msol$
(right) groups. Results are shown for 2PIGG (squares), the bumpy mock
(dotted lines) and the superwind mock (dashed lines).}
\label{fig:dlfb}
\end{figure*}

Having determined the efficiency with which group mass-dependent
galaxy luminosity functions 
are recovered, it is now appropriate to consider 
the real data. The 2PIGG results are shown by the points in
Fig.~\ref{fig:bbandb}. Error bars are the Poisson errors on the number of groups
contributing galaxies to each luminosity bin. In almost all respects,
the real 2PIGG results look very similar to those recovered from the
mock catalogue, shown as dotted lines. The main difference is the
abundance of low luminosity 
($L<L_*/3$) galaxies in higher mass haloes. The mock groups
contain more of these galaxies than are present in high mass 2PIGGs. This
difference is apparent in the panels showing the $M\sim 10^{14}\Msol$
and $M\sim 10^{15}\Msol$ results. The evidence for bumps in the 2PIGG
data is also only tentative.

Fig.~\ref{fig:styb} shows the
$1$, $2$ and $3-\sigma$ contours in the $L_*-\alpha$ plane for the STY
(Sandage, Tammann \& Yahil 1979) fits to the 2PIGG 
galaxy luminosity functions within groups of different mass.
Also shown, without a symbol marking the most likely value, 
are the contours representing the best-fitting parameters for the
whole galaxy population out to $z=0.12$. The parameters for 
the highest mass bin are much less well constrained because there
are fewer total galaxies in these groups than in the more abundant, lower mass
haloes. The 2PIGG sample has a slightly brighter $L_*$ for the $M\sim
10^{14}\Msol$ groups than the $M\sim 10^{13}\Msol$ groups. This trend is
reproduced in the bumpy mock catalogue but not in the superwind mock,
in which the most luminous galaxies no longer reside in the most
massive haloes. 

Fig.~\ref{fig:dlfb} shows how the recovered luminosity
functions differ from the suitably normalised best-fitting Schechter
functions for the 2PIGGs. While the fit works well for both mock and
real data in the $M\sim
10^{13}\Msol$ groups, it becomes an increasingly poor description 
in more massive systems, where an excess of luminous galaxies can
be seen to distort the fit. Also apparent is the overabundance of
low luminosity galaxies in the most massive mock groups relative to
what is found in the 2PIGGs. Finally, the lack of very luminous
galaxies in the superwind mock shows that, despite the imperfections
of the recovery, it is still possible to discriminate between this and the 
bumpy model.

\subsection{$r_{\rm F}$ band results}\label{ssec:glfr}

\begin{figure*}
\centering
\centerline{\epsfxsize=19cm \epsfbox{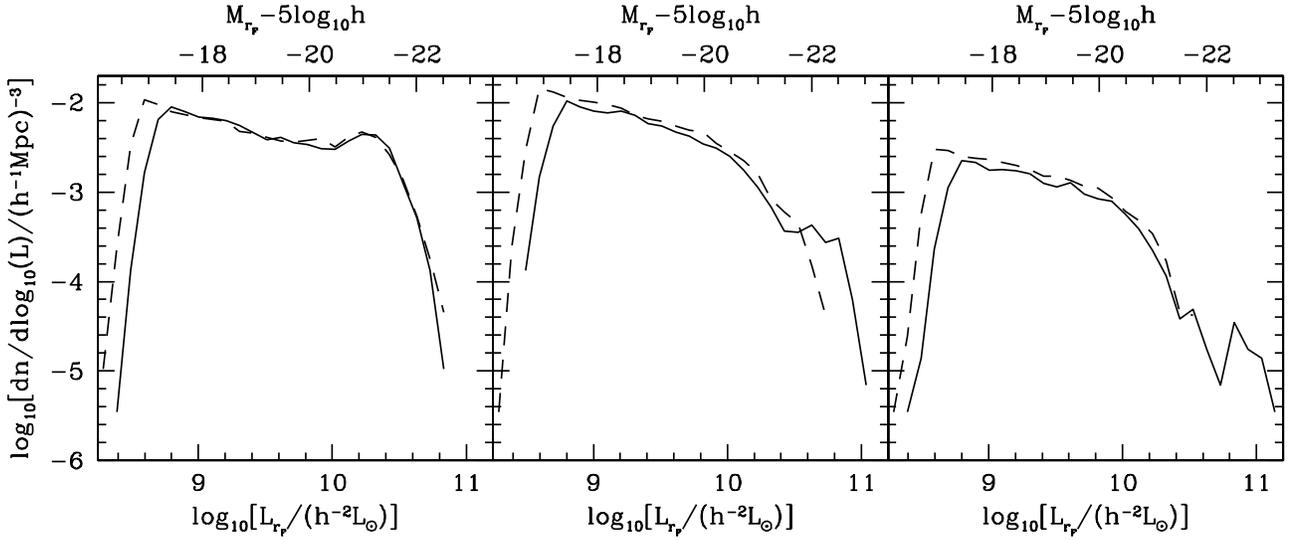}}
\vspace{-11.5cm}
\caption{The $\rf$ band equivalent of Fig.~\ref{fig:bcomp}, showing
the galaxy luminosity functions in groups of different mass in the
two semi-analytical models of galaxy formation. As before, the panels
correspond to groups with $M=10^{13\pm0.5}\Msol$ (left),
$M=10^{14\pm0.5}\Msol$ (centre) and $M=10^{15\pm0.5}\Msol$
(right). Solid lines correspond to the `bumpy' semi-analytical
model, whereas dashed lines correspond to the `superwind' model.}
\label{fig:rcomp}
\end{figure*}

\begin{figure*}
\centering
\centerline{\epsfxsize=19cm \epsfbox{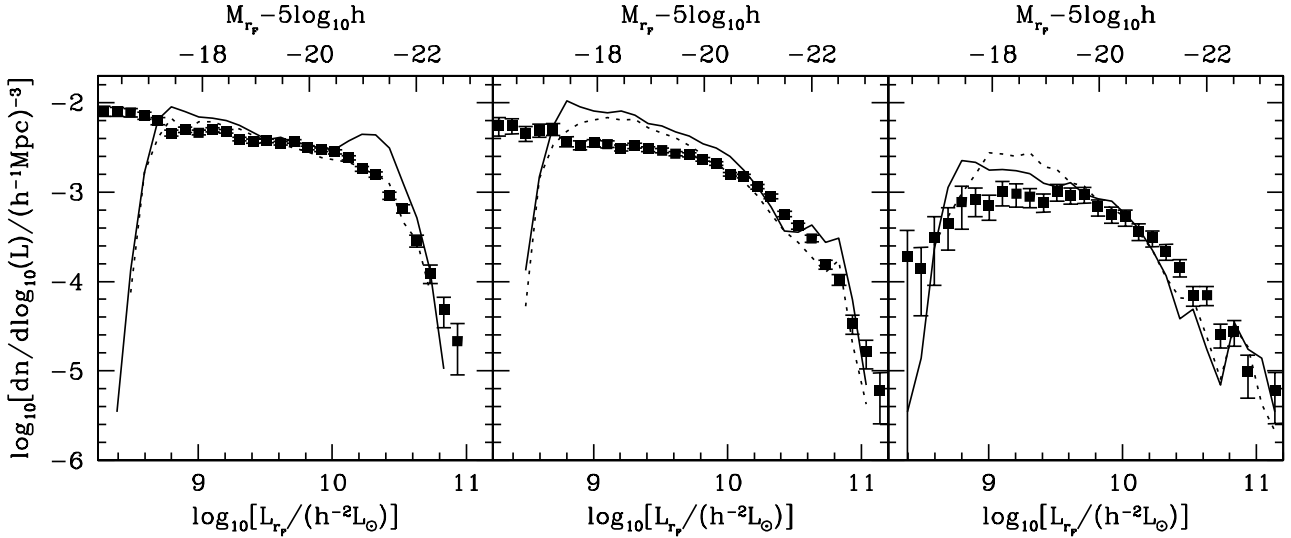}}
\vspace{-11.5cm}
\caption{The $\rf$ band equivalent of Fig.~\ref{fig:bbandb}, showing
both how well the galaxy luminosity functions can be recovered in different
sized groups, and how the mock results compare with the real 2PIGG
data. As before, the group mass ranges are
$M\sim 10^{13}\Msol$ (left), $M\sim 10^{14}\Msol$ (centre) and $M\sim
10^{15}\Msol$ (right). In each case, a solid line shows the
semi-analytical prediction from which the mock catalogues were
constructed, a dotted line shows what is actually measured in the mock
groups, and the filled squares with errors depict the results from the
2PIGG data.}
\label{fig:rbandb}
\end{figure*}

The same analysis performed in Section~\ref{ssec:glfb} for the $\bj$
band data can be performed for the $\rf$ band data. While all galaxies
can be used to calculate the luminosity functions using the $1/V_{\rm
max}$ method, the STY estimation of the best-fitting Schechter
function parameters requires a complete sample, so galaxies are only
included if $14<\rf<b_{\rm J,lim}-1.5$ when determining $L_*$ and
$\alpha$.

\begin{figure}
\centering
\centerline{\epsfxsize=8.4cm \epsfbox{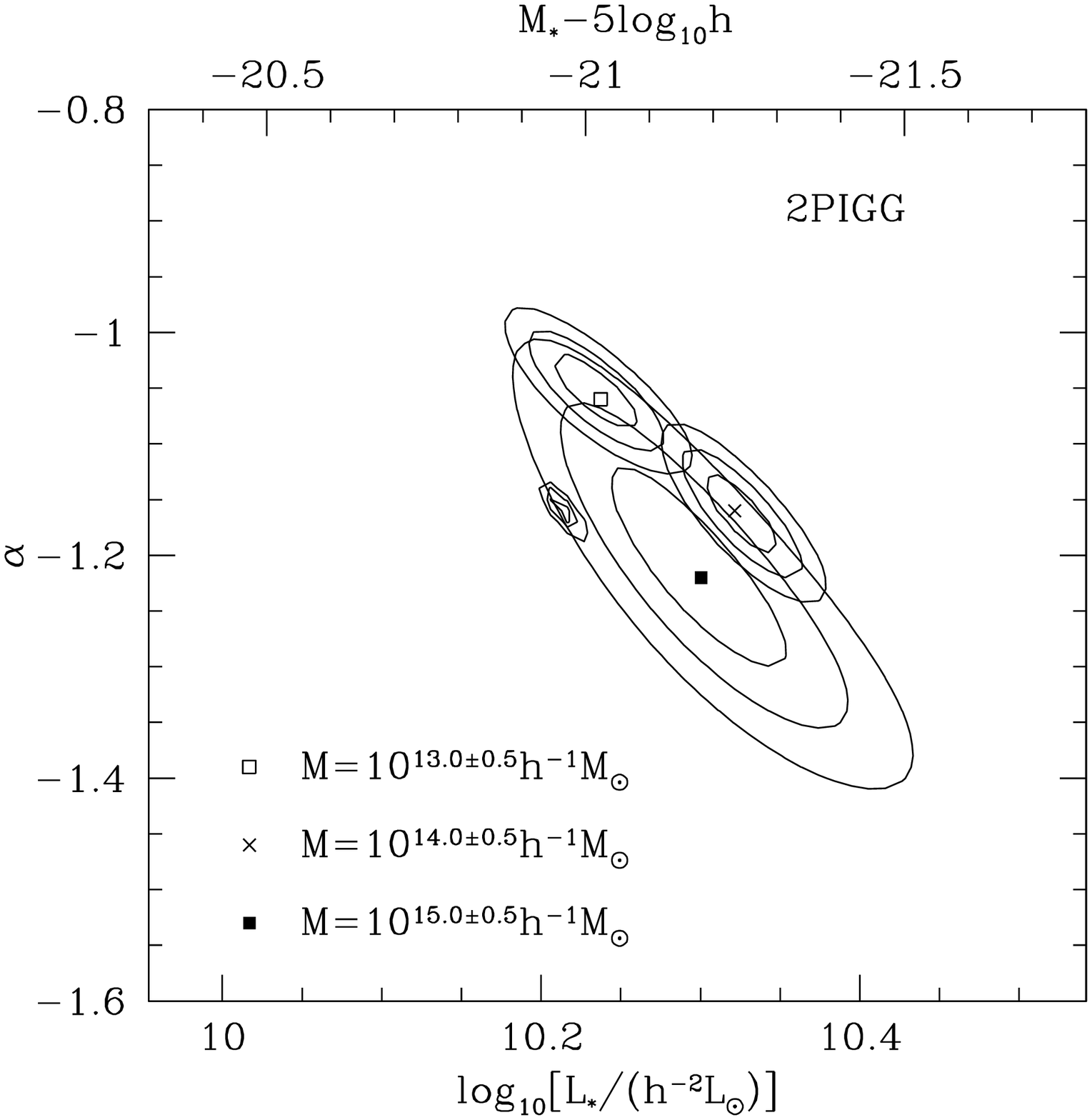}}
\caption{The $\rf$ band equivalent of Fig.~\ref{fig:styb}, showing
the STY-determined most likely Schechter function parameters for
2PIGG galaxy luminosity functions. The different group masses are
represented by the contours enclosing the open square
($M=10^{13\pm0.5}\Msol$), cross ($M=10^{14\pm0.5}\Msol$) and filled
square ($M=10^{15\pm0.5}\Msol$). Results for all galaxies satisfying
$14<\rf<b_{\rm J,lim}-1.5$ and $z<0.12$ are shown by contours with no
central symbol.}
\label{fig:styr}
\end{figure}

\begin{figure*}
\centering
\centerline{\epsfxsize=19cm \epsfbox{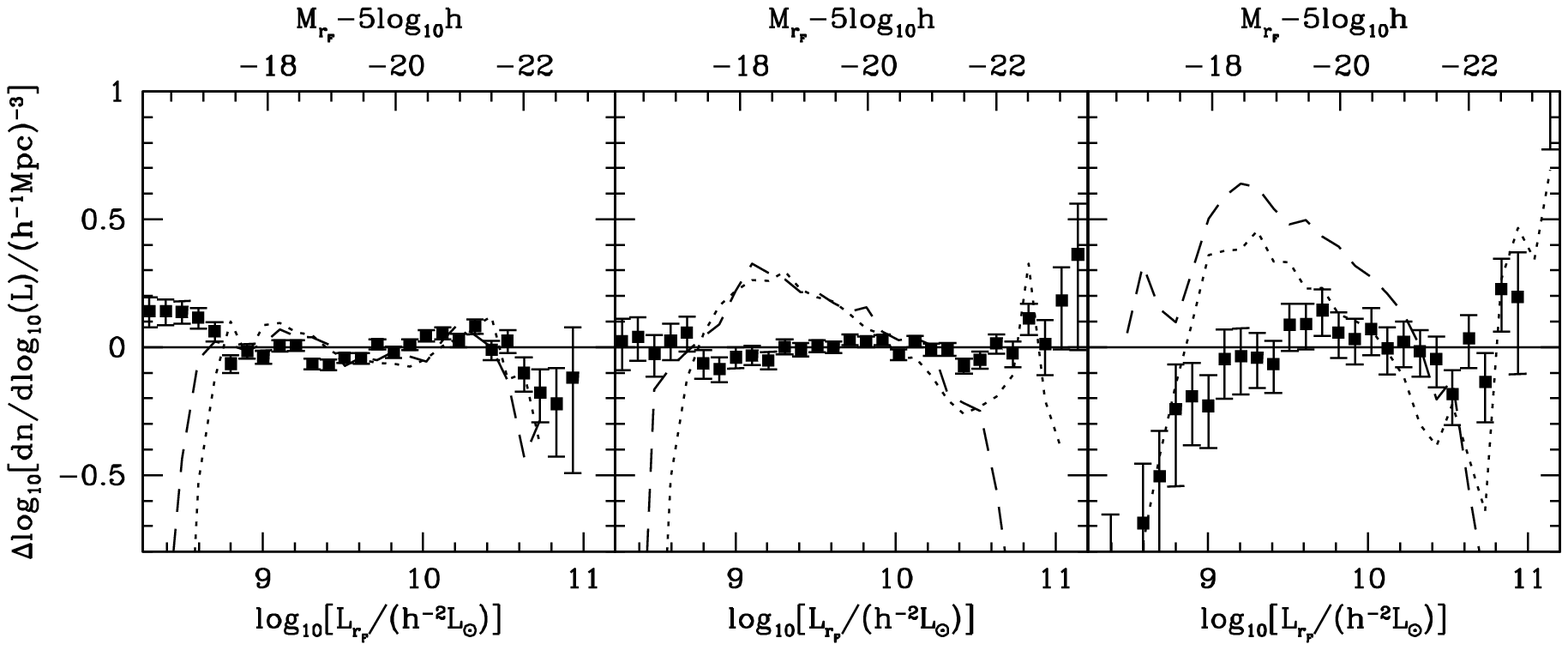}}
\vspace{-11.5cm}
\caption{The $\rf$ band equivalent of Fig.~\ref{fig:dlfb}, showing the
ratios of galaxy luminosity functions in the 2PIGG (squares), bumpy
mock groups (dotted lines) and superwind mock groups (dashed lines)
to the best-fitting Schechter function for the 2PIGGs in each mass
range.  The mass ranges are: $M\sim 10^{13}\Msol$ (left), $M\sim
10^{14}\Msol$ (centre) and $M\sim 10^{15}\Msol$ (right).}
\label{fig:dlfr}
\end{figure*}

Figs~\ref{fig:rcomp},~\ref{fig:rbandb},~\ref{fig:styr}
and~\ref{fig:dlfr} are the 
$\rf$ band equivalents of the previous four figures for the $\bj$
band. Apart from larger error ellipses on the recovered Schechter
function fits, resulting from the smaller number of galaxies being
used to ensure a complete sample, the $\rf$ band results are
qualitatively very similar to those found in the $\bj$ band. The
2PIGGs with $M\sim 10^{14}\Msol$ again have an average
galaxy luminosity function with $L_*$ about $25$ per cent higher than
that for the $M\sim 10^{13}\Msol$ groups, and a slightly steeper faint
end slope. The bumpy mock reproduces this scaling of $L_*$ with halo
mass, but again contains an excess of low luminosity galaxies in the
most 
massive haloes. Once more, the superwind mock exhibits both an excess
of low luminosity galaxies and a deficit of high luminosity galaxies
in both of the higher mass bins, relative to the 2PIGG sample.

\section{Group mass-to-light ratios}\label{sec:mol}

As outlined in the Introduction, the mass-to-light ratio of groups,
$\Upsilon$, contains clues to the nature of galaxy formation, and it
can also be used to estimate the mean mass density of the Universe.
This Section contains a description of the accuracy with which 
$\Upsilon$ can be determined from the 2PIGG catalogue, and addresses 
the dependence of $\Upsilon$ on group luminosity, 
the size of halo in which stars form most efficiently, and the
estimate of the mean mass density, $\Omega_m$.

In trying to recover the typical group mass-to-light ratio for groups of a
particular size, there is no reason to use the entire group
sample. One merely requires an unbiased subset of groups. Experimentation
with the mock catalogues has yielded an appropriate subset, in which 
the bias of the results is minimized. In this scheme,
the smaller groups are only used when they are relatively
nearby and isolated, whereas the restrictions are less stringent for
larger groups. Specifically,
of the $n_{\rm min}=2$ groups, only those with $z<z_{\rm
max}=0.07+0.02[\,\log_{10}(L_{b_{\rm J}}/\Lsol)-10]$ and having 
no neighbouring groups with centres at distances less than 
$d_{\rm min}/\Mpc=2+[\,10-\log_{10}(L_{b_{\rm J}}/\Lsol)]$ were
used to calculate the mass-to-light ratios. 

The reasons for the success of this empirical choice are as follows.
Many of the small groups are fragments cleaved from much bigger groups
by the group-finding algorithm. These typically have large velocity
dispersions, reflecting the size of the halo to which they really
belong. Consequently their mass-to-light ratios are unrepresentatively
high. The nearest neighbour distance restriction is intended to
eliminate these spurious groups from the analysis. The reason for the
redshift limit is the desire that most underlying groups of a
particular size should contain at least two galaxies that will be
detected in a flux-limited survey. The mock catalogues used here are
such that the true mass-to-light ratio of low luminosity groups is
typically lower for groups containing only one detectable
galaxy. Thus, in order to ensure that the low luminosity groups are
representative of the underlying distribution, only the nearby low
luminosity examples are included.

The following two subsections describe the $\bj$ and $\rf$ band
results.  The redder band traces the stellar mass more faithfully than
the blue band, and $\Upsilon_{r_{\rm F}}$ thus provides additional
useful information.

\subsection{$b_{\rm J}$ band results}\label{ssec:molb}

\begin{figure}
\centering
\centerline{\epsfxsize=8.4cm \epsfbox{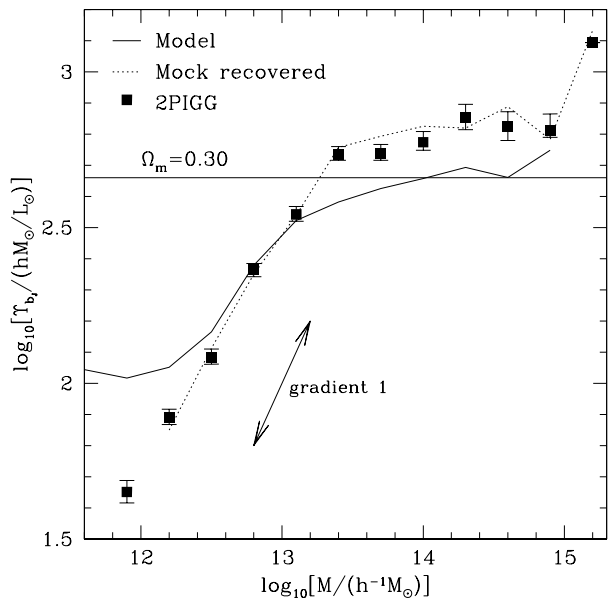}}
\caption{The group mass-to-light ratios as a function of group mass.
The solid line traces the variation of the  median
$\Upsilon_{b_{\rm J}}$ in the semi-analytical model from which the mock
group catalogues were constructed. The dotted line shows the
$\Upsilon_{b_{\rm J}}$  actually recovered from the
mock catalogue using groups with $n_{\rm min}=2$  that lie within a redshift
of $z_{\rm max}=0.07+0.02\,[\,\log_{10}(L_{b_{\rm J}}/\Lsol)-10]$ and have
no neighbouring groups with centres at distances less than 
$d_{\rm min}/\Mpc=2+[\,10-\log_{10}(L_{b_{\rm J}}/\Lsol)]$. The
corresponding measurement from the 2PIGG sample is shown by the filled
squares, with error bars representing the 16th and 84th percentiles
divided by the square root of the number of groups contributing to
each bin. This would be the standard deviation on the
median if the $\Upsilon_{b_{\rm J}}$ values were
distributed Normally in each bin.
The horizontal line indicates the mean mass-to-light ratio of the mock 
universe, 
calculated using the luminosity function of Norberg \etal (2002), and 
the appropriate value of $\Omega_m=0.3$. A line with gradient 1 is
also plotted to illustrate the direction in which errors in the
inferred group masses would move the estimated values.}
\label{fig:molm}
\end{figure}

\begin{figure}
\centering
\centerline{\epsfxsize=8.4cm \epsfbox{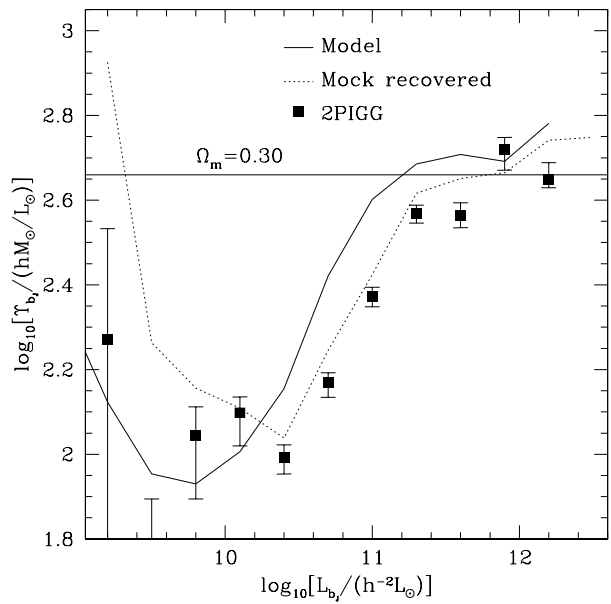}}
\caption{The median mass-to-light ratio of groups as a function of group
luminosity. Line types and symbols are the same as in the previous
figure, as are the values of $n_{\rm min}$, $z_{\rm max}$ and 
$d_{\rm min}$ defining the recovered groups.}
\label{fig:moll}
\end{figure}

Fig.~\ref{fig:molm} shows the dependence of $\Upsilon_{b_{\rm J}}$ on group
mass. The horizontal line indicates the mean mass-to-light ratio in the
mock universe, and the other solid line represents the variation of
the median group $\Upsilon_{b_{\rm J}}$ with halo mass in the
simulation with semi-analytical model galaxies 
from which the mock catalogues were constructed. A dotted line traces what is
actually recovered from the mock catalogue, and the filled squares show the
results from the real 2PIGG subsample of groups. Error bars represent
the 16th and 84th percentiles divided by the square root of the number
of groups in each bin.
As a result of the large spread of accuracies in the
inferred group masses for groups containing only a few members,
there is a strong smearing effect along a line of gradient 
one. For the highest mass groups, for which more galaxies yield better
mass estimates, this trend is reduced, but there is sufficient
contamination from low luminosity groups with greatly overestimated
masses, that the recovered $\Upsilon_{b_{\rm J}}$ is still biased
high by $30-40$ per cent. The main reason why the mock recovered variation
resembles the true behaviour in the parent simulation at intermediate masses is that this
is where the majority of the groups are found (see Fig.~\ref{fig:nm}), 
so there are comparable numbers of groups biasing the results high and
low. In summary, this figure demonstrates that it is unhelpful to
plot correlated variables when they both suffer from the same large
uncertainty. 

A much better method for ordering groups in terms of size is to use
their measured luminosities; these have much
smaller errors than the group masses, 
particularly for groups with only two members (see Fig.~\ref{fig:errorsn}).
Now, instead of mass errors smearing out the results along a line of 
gradient one, there are luminosity uncertainties that act
along a line of gradient minus one. The sign of this
gradient is opposite to the trend of $\Upsilon_{b_{\rm J}}$ with halo
luminosity in the parent simulation, and
the amplitude of this error vector is now substantially reduced.
Fig.~\ref{fig:moll}
shows the corresponding variation of group mass-to-light ratio with
group luminosity. 

In the semi-analytical model, the variation of $\Upsilon_{b_{\rm J}}$
reflects the efficiency of star formation in haloes of different mass.
Star formation is most efficient in haloes with $L \approx 6\times
10^{9} \Lsol$, the point at which the minimum of the solid line
occurs. The dotted line traces the variation of $\Upsilon_{b_{\rm J}}$
recovered from the group mock catalogue. It is
immediately apparent that this recovery, for haloes with $L_{b_{\rm
J}}\gsim10^{10}\Lsol$, is very much better than was the case when group
mass was used to quantify halo size. In particular, both the rise in
$\Upsilon_{b_{\rm J}}$ from small groups to clusters, and the plateau
for the largest clusters are well reproduced. There is a slight bias low, by
$\sim 40$ per cent, during the rise. This can be understood in terms of
the position of the peak in the distribution of groups as a function of
luminosity (Fig.~\ref{fig:nl}), and the typical errors being made,
whereby luminosities are slightly overestimated as a result of
contamination (Fig.~\ref{fig:errorsl}).
%Consequently there is a tendency for groups
%to be pushed into inappropriately high luminosity bins. As their real
%mass-to-light ratios are typical of less luminous groups (\ie lower,
%because of the slope in the mock input $\Upsilon_{b_{\rm J}}$ with
%luminosity), this pulls the measured median $\Upsilon_{b_{\rm J}}$
%down from the correct value.
The recovery of the mass-to-light ratio for groups less luminous than
$L_*$ (for galaxies) is poor. Among the reasons for this is the fact that the
individual redshift measurement errors become comparable with the
typical group velocity dispersions in such small systems. Also, there
is insufficient volume to achieve small statistical errors while
ensuring that these groups contain more than one detectable galaxy.

For the $111$ groups with $\log_{10}[L_{b_{\rm J}}/(\Lsol)]>11.5$ in
the mock catalogue, the recovered median and mean mass-to-light ratios
are $471$ and $549\pm28\, h\MoverL$ respectively. The corresponding
values in the parent simulation are 507 (median) and 520 (mean). Thus,
the mean is overestimated in the mocks by $\sim 6$ per cent, while
the recovered median is $7$ per cent too low. As
the mean is more easily affected by outlying data points
with high measurement errors, in what follows only the median
cluster mass-to-light ratio will be used. The estimate of
$\Omega_m$ from the cluster mass-to-light ratio is based on comparing
this to the mean mass-to-light ratio of the universe. In the parent
simulation, the mean universal value is $458\, h\MoverL$, $10$ per
cent lower than the median recovered mass-to-light ratio of clusters. Thus, for
this mock catalogue at least, the median mass-to-light ratio of the
recovered clusters allows a pretty accurate estimate of the mean mass
density of the simulated universe.

The filled squares with error bars in Fig~\ref{fig:moll}
represent the real 2PIGG mass-to-light
results. These are remarkably similar to what was recovered from the
mock catalogue, showing the same three basic features: nothing useful
at $L\lsim 10^{10}\Lsol$, a rise by a factor of $\sim 5$ up to $L\sim
2 \times 10^{11}\Lsol$, and a constant mass-to-light ratio for the
largest groups. Unlike the case when mass was used to quantify the
group size, this trend of increasing median mass-to-light ratio with larger
groups can no longer be confused with the effect of mass measurement errors. For
the $\log_{10}[L_{b_{\rm J}}/(\Lsol)]>11.5$ groups, of which there are
$96$, the median $\Upsilon_{b_{\rm J}}=429\pm 25\,h\MoverL$, where the 
uncertainty represents the statistical error on the median.
This number is used later to estimate $\Omega_m$.

\begin{figure}
\centering
\centerline{\epsfxsize=8.4cm \epsfbox{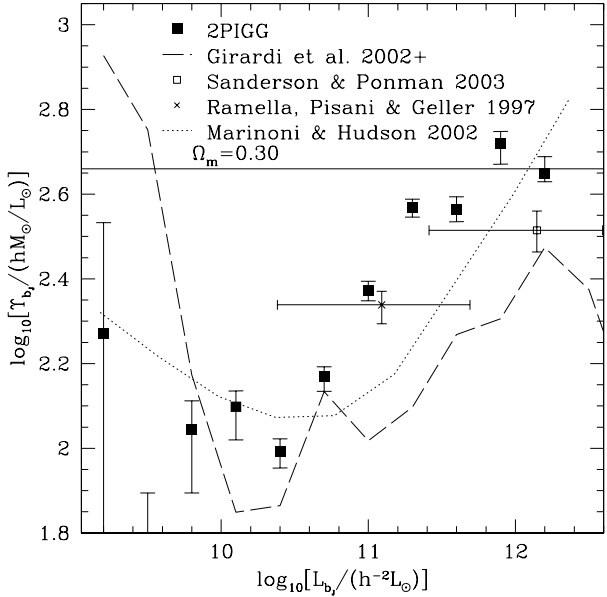}}
\caption{The median mass-to-light ratio of groups as a function of group
luminosity. As in the previous figure, filled squares represent results
for the 2PIGG sample. The horizontal line indicates the mean mass-to-light
ratio determined from the mean luminosity density of Norberg \etal (2001),
assuming $\Omega_m=0.3$. The results of four other studies of B-band group
mass-to-light ratios are also shown, as indicated in the
legend. Vertical error bars on the data points give the uncertainty on
the median $\Upsilon_{b_{\rm J}}$. The horizontal error bars represent
the $16-84$ percentile range of the group luminosities contributing to
these medians.}
\label{fig:moll2}
\end{figure}

\subsection{Comparison with other studies}\label{ssec:molbcomp}

When comparing these results with other studies, it is essential to bear
in mind that there exists a variety of definitions of a group. To
reiterate, the 2PIGG analysis is designed to recover, as well as
possible, groups that resemble those
identified in a $\Lambda$CDM dark matter simulation using the standard
linking length of $0.2$ times the mean interparticle separation. Note
that in Fig.~\ref{fig:moll2}, the raw 2PIGG results are shown for the
purpose of comparison with other studies, despite the fact that they
probably suffer from a small bias of the sort shown in
Fig.~\ref{fig:moll}. In Section~\ref{ssec:molcor}, this bias is
corrected and functions are given which describe the variation of the
corrected median group mass-to-light ratio with group luminosity.

The results of Girardi \etal
(2002; G2002), which extend the database used by Girardi \etal (2000) with
groups mostly found in the Nearby Optical Galaxy (NOG) catalogue, contain
groups from a number of different surveys. Note that these papers both
contain a typo relating to the conversion between $B$ and $\bj$
magnitudes (Girardi, private communication). The correct relation,
for a mean galaxy colour of $(B-V)=0.9$, is
\begin{equation}
B-\bj=0.252.
\label{btobj}
\end{equation}
In conjunction with
\begin{equation}
(B-\bj)_\odot=0.15,
\label{btobjsol}
\end{equation}
this leads to
\begin{equation}
\frac{\left(\frac{L}{L_\odot}\right)_{\bj}}{\left(\frac{L}{L_\odot}\right)_{B}}=1.1,
\label{lbtolbj}
\end{equation}
as was correctly reported in G2002.  Fig.~\ref{fig:moll2} shows the
data from the (CL+PS) sample of G2002, supplemented with 500 other
systems from their set of groups found by applying a percolation
algorithm to the NOG survey (Girardi, private communication).
Relative to the homogenous 2PIGG sample, this set of groups comes from
a number of different sources and, furthermore, they were analyzed
using different types of mass and luminosity estimators. Yet, a
broadly similar behaviour of the mass-to-light ratio is seen.  Care is
needed, however, because the percolation algorithm of G2002 will
presumably yield results that are biased to low mass-to-light ratios,
for the same reason that the 2PIGG results are affected.  Their
percolation algorithm (Giuricin \etal 2000) employs a scaling of the
linking volume with redshift that is different from that used for the
2PIGG catalogue (Eke \etal 2004), so the properties of the recovered
groups will be different.

Sanderson \& Ponman (2003) used a sample of $32$ systems with masses
inferred from their X-ray emission. These groups are concentrated
towards the large mass end of the distribution and show little or no
trend of mass-to-light ratio with group luminosity, comparable with
the 2PIGG results. Again, their results yield slightly lower values of
the mass-to-light ratio than found for 2PIGG. Note that the point
plotted in the figure differs very slightly from that in the Sanderson
\& Ponman paper as a result of the correction of an error in their $B$
to $\bj$ conversion for 3 groups.

Ramella, Pisani \& Geller (1997) used a sample of $406$ groups
identified in the northern CfA2 survey, with masses estimated from
optical data. While the median mass-to-light ratio is lower than that
of the Sanderson \& Ponman groups, the CfA2 groups are typically
smaller and the 2PIGG results indicate that $\Upsilon_{b_{\rm J}}$
should be smaller for such groups. However, once more, the
mass-to-light ratio in this sample is lower than that inferred from
the 2PIGG catalogue.

Marinoni \& Hudson (2002) have indirectly inferred a mass-to-light ratio
variation for groups by finding the mapping between the 
Press-Schechter (1974) mass function, for some assumed
cosmological model, and the luminosity function
of groups measured from the NOG sample by Marinoni \etal (2002).
Their results are given in the $B$ band, and have been
converted to the $\bj$ band using equation~(\ref{lbtolbj}).
This curve shows a similar trend to the 2PIGG results but, again, with 
an offset. In general, estimators of the luminosity function of groups
tend to be biased towards high luminosities 
as a result of the inevitable inclusion of interlopers in the groups, 
the statistical errors in the inferred group luminosities 
and the steep decline in the abundance of objects with increasing
size. This would lead to an overestimate of the luminosity at a
particular abundance (or mass), biasing low 
the inferred mass-to-light ratio.  Indeed, estimating the luminosity
function of groups in mock 2PIGG catalogues suggests that this
bias could plausibly yield an overestimate of the group luminosity 
by a factor of at least 2 at a particular abundance (Eke
\etal in preparation). If
a similar size of bias were present in the sample used by Marinoni \&
Hudson, then this would account for their significantly lower
mass-to-light ratios at $L\sim 10^{11}\Lsol$.

One factor that would bias the 2PIGG mass-to-light values slightly
high is the incompleteness in the parent 2dFGRS catalogue (Pimbblet
\etal 2001; Norberg \etal 2002; Cross \etal 2004). Such an effect,
however, is small compared to the difference between the 2PIGG results
and those of the other studies shown in Fig.~\ref{fig:moll2}.  The
most likely other reason why the 2PIGG results could be systematically
incorrect is if the mass estimator employed here is not
appropriate in the real world, as discussed in
Section~\ref{sec:data}. This would happen if galaxies traced
real dark matter haloes in a different way from what was assumed in
making the mock catalogues, either because they had a different
velocity dispersion to the underlying dark matter, or because they had
a different projected spatial distribution (De Lucia \etal 2004;
Diemand, Moore \& Stadel 2004). In order for biases of
this kind to give rise to the observed variation of the mass-to-light
ratio with group size, however, they would need to depend very
strongly on group size.  It is difficult to imagine that such biases
could give rise to the factor of $\sim 5$ variation seen in the data.

\subsection{$\rf$ band results}\label{ssec:molr}

\begin{figure}
\centering
\centerline{\epsfxsize=8.4cm \epsfbox{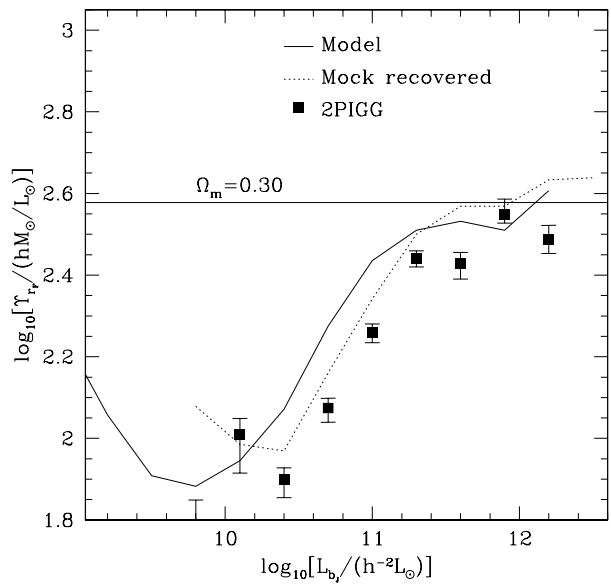}}
\caption{The variation of the $r$ band group mass-to-light ratio
with group $\bj$ luminosity.  A solid line traces the median behaviour
in the parent simulation, and the dotted line shows what is actually
recovered in the mock catalogue. Filled squares represent the results
from the 2PIGG catalogue, using the same redshift and minimum group
separation as were applied to the $\bj$ band data.}
\label{fig:molfinalr}
\end{figure}

\begin{figure}
\centering
\centerline{\epsfxsize=8.4cm \epsfbox{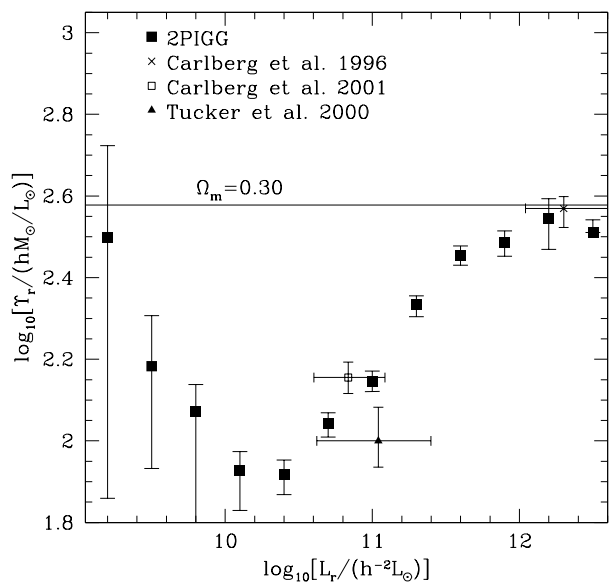}}
\caption{The variation of the $\rf$ band group mass-to-light ratio
with group luminosity. Filled squares represent the results from the
2PIGG catalogue, and the other symbols show other observational
determinations, as detailed in the legend. Note that the Carlberg
\etal results have luminosities measured in the Gunn $r$ band, and the
Tucker \etal data are in the LCRS $R$ band.}
\label{fig:molrealr}
\end{figure}

As described in Appendix~A, the 2dFGRS combined with the
SuperCOSMOS $\rf$ data provides a well defined sample of galaxies with
$14<\rf<b_{\rm J,lim}-1.5$. While applying these flux limits reduces the
number of galaxies available to calculate the group luminosity, the
missing faint galaxies contribute little to the total,
so the accuracy with which group $\rf$ band
luminosities can be inferred is comparable to that for the
$\bj$ band, as was shown in Section~\ref{ssec:accu}. 

Fig.~\ref{fig:molfinalr} shows how well the variation of
$\Upsilon_{r_{\rm F}}$ in the parent simulation can be recovered in
the mock catalogue, as a function of group $\bj$ luminosity. (This is
a slightly better determined measure of 2PIGG size than the $\rf$
luminosity.) The results are remarkably similar to those shown in
Fig.~\ref{fig:moll} for $\Upsilon_{b_{\rm J}}$. Firstly, the intrinsic
variation present in the parent simulation is quite accurately
recovered in the mocks. Secondly, as was the case for
$\Upsilon_{b_{\rm J}}$, the 2PIGG data agree remarkably well with the
behaviour seen in the mock catalogue.  The main difference between the
$\rf$ and $\bj$ results is the size of the change in typical
mass-to-light ratio over the reliably probed range of group
luminosities. This is not as extreme in the $\rf$ band, a factor of
$\sim 3.5$ rather than the factor of $\sim 5$ seen in the $\bj$ band --
as one would expect if halo size has a stronger effect on recent star
formation than on the overall stellar mass.

Fig.~\ref{fig:molrealr} compares the 2PIGG results with those from
CNOC groups and clusters (Carlberg \etal 1996, 2001) and the LCRS
groups (Tucker \etal 2000). Unlike in the $\bj$ band case, there is
now some agreement between the 2PIGGs and some other work. Note that
the Carlberg \etal data points have luminosities measured in the Gunn
$r$ band and include a correction to redshift zero, whereas the Tucker
\etal data are measured in the LCRS $R$ band. Once again, the
horizontal error bars illustrate the $16-84$th percentile range in group
luminosity contributing to each point.

\begin{figure}
\centering
\centerline{\epsfxsize=8.4cm \epsfbox{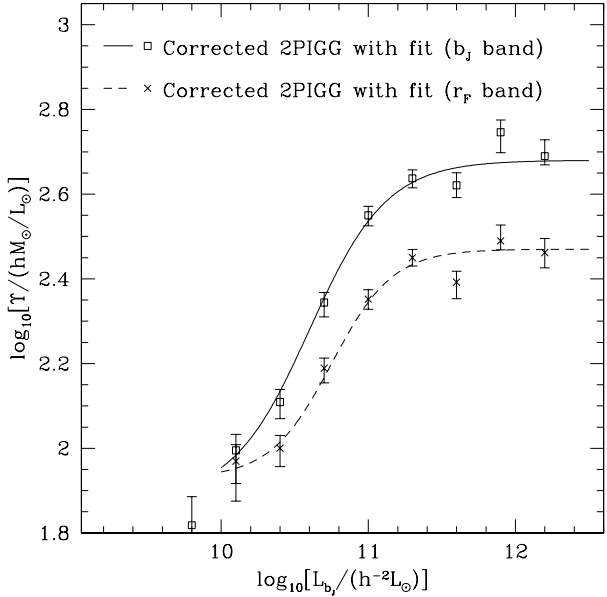}}
\caption{The corrected variation of the 2PIGG group mass-to-light ratio
with group luminosity, in both the $\bj$ and $\rf$
bands. Equations~(\ref{fitmol}) for $\Upsilon_{b_{\rm J}}$
and~(\ref{fitmolr}) for $\Upsilon_{r_{\rm F}}$ are shown as solid
and dashed lines. The corresponding 2PIGG data are shown with open squares
and crosses.}
\label{fig:molcor}
\end{figure}

\subsection{Corrected mass-to-light results}\label{ssec:molcor}

It was shown in Sections~\ref{ssec:molb} and~\ref{ssec:molr} that the
values of $\Upsilon$ as a function of $L_{bj}$ recovered in the mock
catalogues are slightly biased low relative to the values in the
parent simulation. This bias is reproducible in mocks
constructed from other simulations, for example, from the less
clustered mass distribution of a $\Lambda$CDM simulation with
$\sigma_8\approx0.7$ analyzed at $z\approx 0.1$ (rather than the
default simulation with $\sigma_8=0.9$ and $z=0.1$).  

It seems plausible that a bias of this kind is also present in the
2PIGG sample results. A correction factor is defined as the
ratio between the luminosity-dependent values of $\Upsilon$
measured in the parent simulation and in the mocks (in both the $\bj$
and $\rf$ bands). The 2PIGG results are simply multiplied by this
factor to obtain an estimate of the true underlying 
group mass-to-light ratio variation in the real Universe. The
corrected values are shown with open squares ($\bj$ band) and crosses
($\rf$ band) in Fig.~\ref{fig:molcor}. The error bars on the data
points still represent the size of the statistical uncertainty on the
median. Approximate fits to the corrected 2PIGG median mass-to-light
ratios are shown by the solid ($\bj$) and dashed ($\rf$) lines, which
have the following functional forms:
\begin{equation}
\eqalign{
\log_{10}\Upsilon_{b_{\rm J}} &=2.28 \cr
&+0.4\,{\rm tanh}\{1.9\,[\log_{10}(L_{b_{\rm
J}})-10.6]\} \cr
}
\label{fitmol}
\end{equation}
and
\begin{equation}
\eqalign{
\log_{10}\Upsilon_{r_{\rm F}} &=2.20 \cr
&+0.27\,{\rm tanh}\{2.4\,[\log_{10}(L_{b_{\rm J}})-10.75]\}. \cr
}
\label{fitmolr}
\end{equation}
These equations are valid for $10\le\log_{10}(L_{b_{\rm
J}}/\Lsol)\le12.5$. Since the 2PIGG results are so similar to those
recovered from the mock catalogue, these fits to the underlying
`truth' in the real Universe also provide a decent description of
the behaviour in the parent simulation, as traced by the solid lines
in Figs~\ref{fig:moll} and~\ref{fig:molfinalr}.  The correction is
sufficiently small that all the basic features of the results discussed
above remain.

The correction changes the median mass-to-light ratio of the $98$ clusters
with $\log_{10}[L_{b_{\rm J}}/(\Lsol)]>11.5$ from $\Upsilon_{b_{\rm
J}}=427\pm 24\,h\MoverL$ to $466 \pm 26 \,h\MoverL$. Together with the
global mean $\bj$ band luminosity density and its uncertainty,
inferred from the luminosity function measured by Norberg \etal
(2002), this implies\footnote{Assuming that the galaxies trace the
mass in clusters, as is customary when using this method} that
$\Omega_m=0.31\pm 0.03$. If, as was the case in the mocks, the median
mass-to-light ratio of the largest clusters overestimates the mean
mass-to-light ratio of the Universe by $\sim 11$ per cent, then this
estimate should come down accordingly to $\Omega_m=0.28\pm 0.03$. This
uncertainty only includes the statistical error.

Possible sources of systematic uncertainty, in addition to the two
model-dependent corrections already applied to the raw 2PIGG results,
would come from systematic errors in the estimation of the cluster
mass-to-light ratios. These could arise from the underestimate of the
cluster luminosities resulting from incompleteness in the parent
2dFGRS catalogue. In fact, Norberg
\etal (2002) do assume a $9$ per cent incompleteness in the 2dFGRS
parent catalogue when computing the mean $\bj$-band luminosity density
of the Universe, so for consistency the 2PIGG luminosities should be
similarly increased. The resulting lower mass-to-light ratios yield
$\Omega_m=0.26\pm 0.03$. Another, less readily quantified, potential
systematic uncertainty would come from a bias in the
mass estimator if real galaxies do not trace dark matter haloes in the
way assumed in the mocks. As discussed in Section~\ref{sec:data} this
could plausible be a few tens of per cent.

It is interesting that both this estimate of $\Omega_m$ and its
statistical uncertainty are very similar to the values quoted by
Spergel \etal (2003) from the combination of WMAP microwave background
data and the 2dFGRS galaxy power spectrum. This agreement, in itself,
suggests that the systematic uncertainties in either estimate are
likely to be small and provides a welcome consistency check of the
entire paradigm of structure formation by hierachical clustering from
CDM initial conditions. Further reassurance can be gained from using
the $\rf$-band mass-to-light ratios of the same set of the most
luminous clusters and the $\rf$-band luminosity density of the
Universe (see Appendix~A). This yields an estimate of
$\Omega_m=0.25\pm0.04$, which is consistent with the value inferred
from the $\bj$-band data.

\section{Conclusions}\label{sec:conc}

This paper presents an analysis of the galactic content of the 2PIGG
catalogue of groups and clusters identified in the 2dFGRS,
focusing on the galaxy luminosity function and total mass-to-light
ratio in groups of different size in both the $\bj$ and $\rf$ bands. In
constructing the 2PIGG catalogue (Eke \etal 2004) and in the present
analysis, a well defined methodology has been followed based on the
extensive use of mock catalogues. The starting points are N-body
simulations of the evolution of dark matter in a $\Lambda$CDM universe in
which model galaxies are added with properties calculated according to
a semi-analytical model of galaxy formation based on the precepts laid
out in Cole \etal (2000). The parent simulation provides a full
description of the model galaxy distribution, free from the
distortions inevitably introduced by observational procedures, such as
selection effects, observational errors, etc. These
distortions are modelled carefully in order to generate mock catalogues that
correspond to artificial 2dFGRSs.
Since the properties of the parent simulation are 
known, the mock catalogues provide a rigorous way to quantify the
systematic uncertainties in the derived properties of interest 
as well as a good guide to possible systematic errors. Of
course, the mock catalogues also enable a detailed comparison between the
real data and the model from which the mocks were generated.

The main results of this paper are: 

\noindent
(1) The galaxy luminosity function is different in groups of
different mass. These luminosity functions are moderately well
described by Schechter functions for which, as the group mass
increases, the characteristic luminosity, $L_*$, increases and the
faint end slope, $\alpha$, becomes more negative. However, the galaxy
luminosity functions in the groups found in the original simulations
are not well described by Schechter functions. For example, they
exhibit a `bump' at the bright end whose amplitude varies with group
mass reflecting the relative importance of the central galaxy and the
satellites in the groups. Hints of bumps are also found in the largest 2PIGG
clusters.\\
(2) The median group mass-to-light ratio, $\Upsilon$, also varies
with halo size (which is most robustly characterized by total $\bj$
halo luminosity). The mock catalogues indicate that, in the 2PIGG
catalogue, $\Upsilon$ is reliably determined for groups of size
ranging from that of the Local Group to the richest clusters in the
survey. Over this range, which extends from $L_{b_{\rm
J}}=10^{10}\Lsol$ to $L_{b_{\rm J}}= 10^{12}\Lsol$, $\Upsilon_{b_{\rm
J}}$ increases by a factor of $5$ whereas $\Upsilon_{r_{\rm F}}$
increases by a factor of $3.5$. At the highest luminosities, $\Upsilon$
becomes roughly constant in both bands. The semi-analytical models
predict an upturn in $\Upsilon$ at luminosities lower than $L_{b_{\rm
J}}\sim 10^{10}\Lsol$ at which galaxy formation is most
efficient. Unfortunately, the 2PIGG catalogue does not contain enough small
groups with sufficiently accurate estimates of $\Upsilon$ to locate
the theoretically expected minimum.\\
(3) For the $L_{b_{\rm J}}>3\times 10^{11}\Lsol$ objects, the median
mass-to-light ratio is $\Upsilon_{b_{\rm J}}= 466 \pm 26$
(statistical) $h \MoverL$, independently of
cluster size. Assuming that this value reflects the cosmic mean (\ie
that $\Upsilon$ in these richest clusters has converged to the global
value) allows a determination of the mean cosmic density,
$\Omega_m$. Adopting the $\bj$ luminosity density inferred from the
2dFGRS galaxy luminosity function by Norberg \etal (2002), leads to
$\Omega_m=0.26\pm 0.03$. (This estimate includes two small
corrections derived from the mock catalogues as discussed in
Section~\ref{ssec:molcor}, and a third small correction for
incompleteness in the 2dFGRS parent catalogue.) 
This value and its uncertainty are in
excellent agreement with the values inferred by Spergel \etal (2003)
from a combination of microwave background data and the galaxy power
spectrum in the 2dFGRS.

The agreement of the 2PIGG results with the predictions of the
simulations is impressive, and provides confidence in the basic
picture of galaxy groups tracing virialized haloes of dark
matter.  The simulations suggest two fruitful avenues for improving
upon this work: (i) probe smaller groups to test whether the $M/L$
ratio does indeed pass through the expected minimum around a group
mass of $10^{12} \Msol$; (ii) create catalogues with better resolved
groups to see if the non-Schechter nature of the theoretically
predicted luminosity functions can be revealed. Both these targets
will require new large redshift surveys that probe substantially
further down the luminosity function than is possible with the current
dataset.

\section*{ACKNOWLEDGEMENTS}

The 2dF Galaxy Redshift Survey was
made possible through the dedicated efforts of the staff of the
Anglo-Australian Observatory, both in creating the 2dF instrument and
in supporting it on the telescope.
VRE and CMB are Royal Society University Research Fellows. PN is a
Zwicky Fellow. JAP is a PPARC Senior Research Fellow.

\begin{appendix}

\section{Calculating group $\rf$ luminosities}\label{app:lr}

\begin{figure}
\centering
\centerline{\epsfxsize=8.4cm \epsfbox{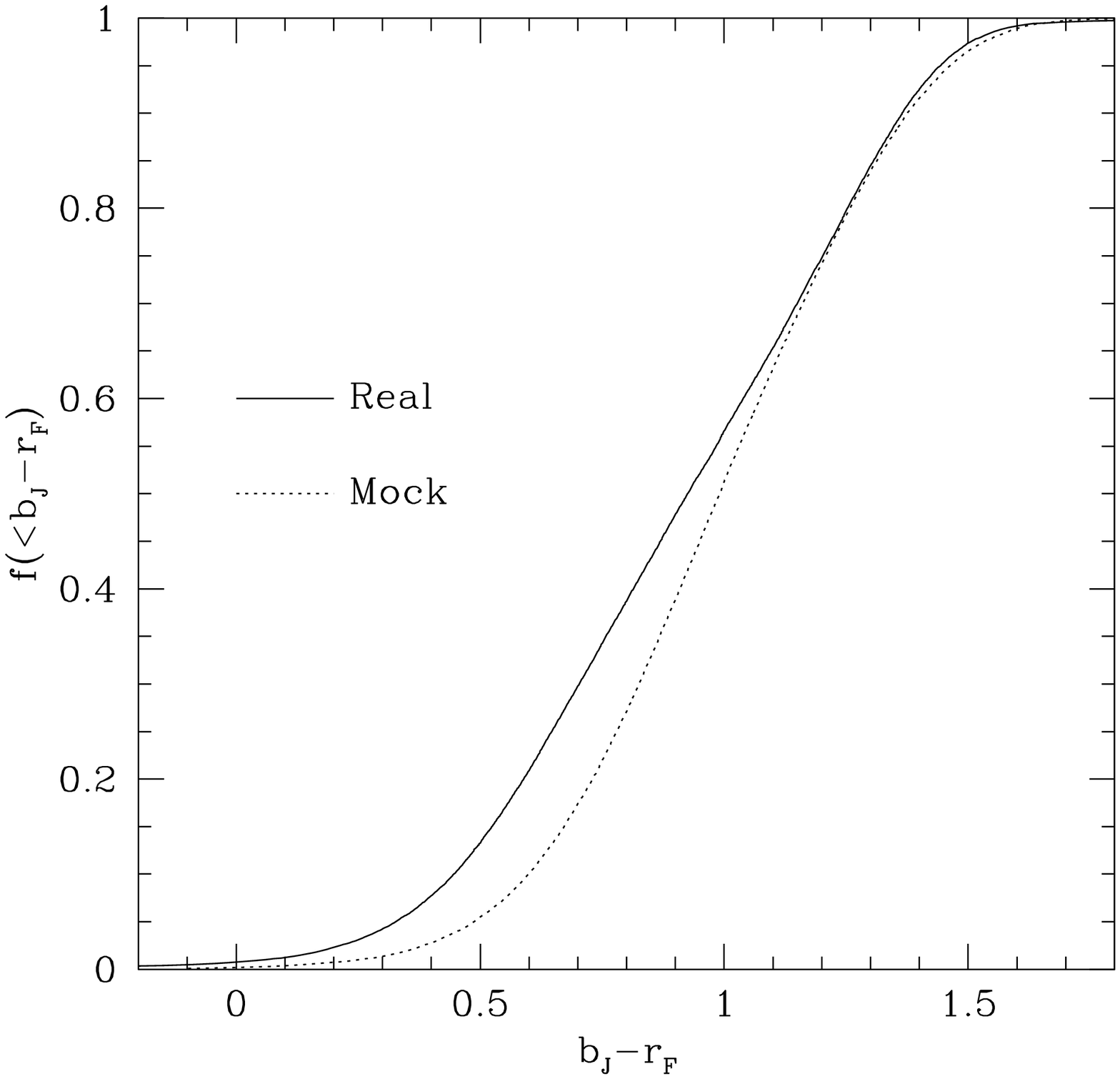}}
\caption{The cumulative distribution of galaxy $\bj-\rf$ values for
all $z<0.12$ galaxies in the 2dFGRS (solid line) and mock (dotted
line) catalogues.}
\label{fig:appcol}
\end{figure}

As the 2dFGRS is a $\bj$ selected survey, extra consideration is
required in order to include the SuperCOSMOS $\rf$ band data in 
the analysis. For instance, it is necessary to determine 
an appropriate local flux limit to which the $\rf$ band data are
complete. The term complete is used here without reference to the
well-quantified incompleteness in the redshifts measured for objects
in the 2dFGRS parent catalogue, and the possible incompleteness in the
parent catalogue itself. A local $\rf$ band flux limit is determined
from the distribution of observed galaxy colours. Fig~\ref{fig:appcol}
shows the cumulative distribution of the galaxy $\bj-\rf$ values for
the sets of $z<0.12$ galaxies in both the mock and the 2dFGRS data.
As at least $95$ per cent of galaxies are bluer than $\bj-\rf=1.5$,
the local minimum $\rf$ band flux limit is set equal to 
$r_{\rm F,lim}=b_{\rm J,lim}-1.5$. From this figure, one can also
infer that the bright $\bj>14$ limit can essentially be transferred
into the $\rf$ band, such that the $\rf$ band sample is also complete
for $\rf>14$. 
Thus, a `complete' $\rf$-limited sample can be made to a locally
defined flux limit corresponding to $14<\rf<b_{\rm J,lim}-1.5$. 
This lower flux limit removes about $29$ per cent of the galaxies from
the sample, whereas only $\sim 0.4$ per cent were discarded for being
brighter than the upper flux limit.

Having found flux limits defining a complete $\rf$ band sample,
the intention is, as before, to tot up the galaxy
luminosities taking into account the weights associated with the redshift
incompleteness, and then apply the appropriate Schechter function correction
for the galaxies that are too faint to make the flux cut. However, the
galaxies included above the $\rf$ flux limit are now a bright 
subsample of the 2dFGRS parent catalogue. Thus, they are no longer
drawn from the same population as the subsample of parent catalogue 
galaxies without measured
redshifts, and so equation~(\ref{easyw}) is no longer an unbiased way to
correct for this redshift incompleteness. This complication was
sidestepped by redistributing only the weights of
galaxies without redshifts and with $14<\rf<b_{\rm J,lim}-1.5$.
Thus, the subsample of galaxies
with redshifts and sufficiently high $\rf$ band fluxes is now a 
similar population to those galaxies without redshifts whose weights
are redistributed.
As only about $2/3$ of the 2dFGRS galaxies satisfy the
$\rf$ flux limits, the weights were redistributed to the nearest $7$
projected galaxies only. This should match the angular smoothing scale
with that used in the $\bj$ band.

\begin{figure}
\centering
\centerline{\epsfxsize=8.4cm \epsfbox{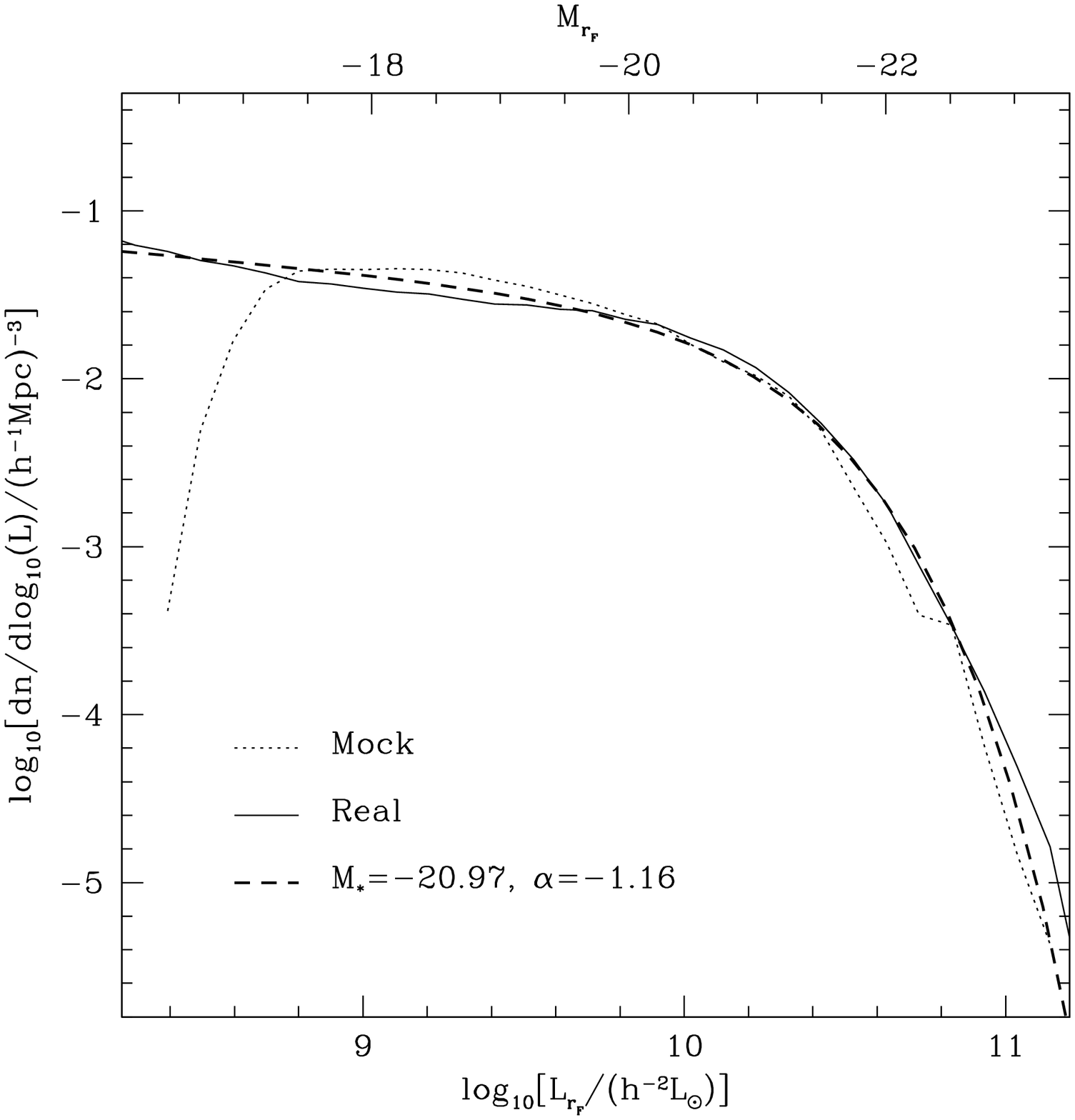}}
\caption{The $\rf$ band luminosity functions, estimated using a
$1/V_{\rm max}$ method, for all $z<0.12$ galaxies in the mock (dotted
line) and 2dFGRS (solid line). The dashed line shows the suitably normalised
Schechter function resulting from an STY fit to the 2dFGRS data.}
\label{fig:apprlf}
\end{figure}

In addition to the
correction for redshift incompletion, galaxies below the flux limit
also need to be accounted for in determining the total group
luminosity. This was accomplished by assuming that the galaxy luminosity
function in the groups can be fitted by a Schechter function with
$(M_*, \alpha)=(-20.97,-1.16)$. This is the STY-estimated Schechter
function for the $\rf$ band galaxy luminosity function of all $z<0.12$
galaxies in the 2dFGRS. In calculating this, the following
$k+e$-correction, derived using Bruzual \& Charlot (1993) models, was used:
\begin{equation}
k+e=\frac{z+2z^2}{1+9z^{2.5}}.
\label{kpluser}
\end{equation}
Choosing a normalisation to match the luminosity function amplitude
inferred using a $1/V_{\rm max}$ estimator leads to $\phi_*=(1.2 \pm
0.1)\times 10^{-2}({\it h}^{-1}\, {\rm Mpc})^{-3}$. This value, which
neglects the incompleteness in the parent catalogue of the 2dFGRS,
gives a mean $\rf$ luminosity density in the Universe of
$\rho_{r_{\rm F}}\approx 2.2\times 10^{-2}\,{\it h}{\rm L_\odot}{\rm Mpc}^{-3}$.
As can be seen in Fig.~\ref{fig:apprlf}, the $\rf$ band luminosity
functions estimated using the $1/V_{\rm max}$ method are similar in
the 2dFGRS and the mock catalogue (which yields $M_*=-20.79$ and
$\alpha=-1.17$). For the $\rf$ band, $M_\odot=4.57$.

\end{appendix}

\end{document}